\documentclass{aa}  

\usepackage{graphicx}
\usepackage[inline]{enumitem} 
\usepackage{hyperref}
\begin{document} 

   \title{Kinematic and metallicity properties of the Aquarius dwarf galaxy from FORS2 MXU spectroscopy.\thanks{Based on observations made
          with ESO telescopes at the La Silla Paranal Observatory as part of the 
          program 091.B-0331.}}
   \author{L. Hermosa Mu\~{n}oz \inst{1,2,3}
        \and
        S. Taibi\inst{2,3}
        \and
        G. Battaglia\inst{2,3}
        \and
        G. Iorio\inst{4}
        \and
        M. Rejkuba\inst{5}
        \and
        R. Leaman\inst{6}
        \and
        A. A. Cole\inst{7}
        \and
        M. Irwin\inst{4}
        \and
        P. Jablonka\inst{8,9}
        \and
        N. Kacharov\inst{6}
        \and
        A. McConnachie\inst{10}
        \and
        E. Starkenburg\inst{11}
        \and
        E. Tolstoy\inst{12}
          }

   \institute{
        Instituto de Astrof\'isica de Andaluc\'ia - CSIC, Glorieta de la Astronom\'ia s/n, 18008 Granada, Spain \\
              \email{lhermosa@iaa.es}
        \and 
        Instituto de Astrof\'isica de Canarias, C/ V\'ia L\'actea s/n, 38205 La Laguna, Tenerife, Spain
        \and 
        Departamento de Astrof\'isica, Universidad de La Laguna, 38205 La Laguna, Tenerife, Spain
        \and  
        Institute of Astronomy, University of Cambridge, Madingley Road, Cambridge CB3 0HA, UK
        \and
        European Southern Observatory, Karl-Schwarzschild Strasse 2, D-85748 Garching, Germany
        \and
        Max Planck Institute for Astronomy, K\"{o}nigstuhl 17, D-69117 Heidelberg, Germany
        \and
        School of Natural Sciences, University of Tasmania, Private Bag 37 Hobart, Tasmania 7001, Australia
        \and
        Institute of Physics, Laboratory of Astrophysics, École Polytechnique Fédérale de Lausanne (EPFL), 1290 Sauverny, Switzerland
        \and
        GEPI, CNRS UMR 8111, Observatoire de Paris, PSL Research University, F-92125, Meudon, Cedex, France
        \and
        National Research Council, Herzberg Institute of Astrophysics, 5071 West Saanich Road, Victoria, BC V9E 2E7, Canada
        \and
        Leibniz-Institut f\"{u}r Astrophysik Potsdam (AIP), An der Sternwarte 16, 14482 Potsdam, Germany
        \and
        Kapteyn Astronomical Institute, University of Groningen, 9700AV Groningen, The Netherlands
             }

   \date{Received M DD, YYYY; accepted M DD, YYYY}

 
  \abstract{
   Dwarf galaxies found in isolation in the Local Group (LG) are unlikely to have interacted with the large LG spirals, and therefore environmental effects such as tidal and ram-pressure stripping should not be the main drivers of their evolution.}
   {We aim to provide insight into the internal mechanisms shaping LG dwarf galaxies by increasing our knowledge of the internal properties of isolated systems. Here we focus on the evolved stellar component of the Aquarius dwarf galaxy, whose kinematic and metallicity properties have only recently started to be explored.}
   {Spectroscopic data in the region of the near-infrared Ca~II triplet lines has been obtained with FORS2 at the Very Large Telescope for 53 red giant branch (RGB) stars. These data are used to derive line-of-sight velocities and [Fe/H] of the individual RGB stars.}
   {We have derived a systemic velocity of $-142.2^{+1.8}_{-1.8}$ km s$^{-1}$, in agreement with previous determinations from both the HI gas and stars. The internal kinematics of Aquarius appears to be best modelled by a combination of random motions (l.o.s. velocity dispersion of $10.3^{+1.6}_{-1.3}$\,km\,s$^{-1}$) and linear rotation (with a gradient $-5.0^{+1.6}_{-1.9}$\,km\,s$^{-1}$\,arcmin$^{-1}$) along a P.A. $= 139_{-27}^{+17}$\,deg, broadly consistent with the optical projected major axis. This rotation signal is significantly misaligned or even counter-rotating to that derived from the HI gas. We also find the tentative presence of a mild negative metallicity gradient and indications that the metal-rich stars have a colder velocity dispersion than the metal-poor ones.}
   {This work represents a significant improvement with respect to previous measurements of the RGB stars of Aquarius, as it doubles the number of member stars already studied in the literature. We speculate that the misaligned rotation between the HI gas and evolved stellar component might have been the result of recent accretion of HI gas, or re-accretion after gas-loss due to internal stellar feedback.}

   \keywords{Techniques: spectroscopic - Galaxies: kinematics and dynamics - Local Group - Galaxies: dwarf - Galaxies: abundances - Galaxies: stellar content}
 
   \titlerunning{Kinematic and metallicity properties of the Aquarius dwarf galaxy.}
   \authorrunning{L. Hermosa Mu\~noz et al.}
   \maketitle
%

\section{Introduction}
\label{sec:intro}

Dwarf galaxies are objects of great interest for galaxy formation and evolution studies because they are the smallest and most numerous galaxies in the Universe. The Local Group (LG) hosts a large number of these systems, which can be studied in great detail: it is possible to gather information on the properties of their stellar component over most of the lifetime of the galaxies by studying individual low-mass stars, such as red giant branch (RGB) stars. 
    
 The internal kinematic and metallicity properties of the "classical" (pre-SDSS) dwarf galaxies orbiting around the Milky Way (MW) have been well-studied due to the favourable combination of close distance, luminosity and angular size, that makes a perfect match for existing wide-area multi-object spectrographs. However, dwarf galaxies have relatively small masses and this arguably makes their evolution susceptible both to internal and external effects. The study of the LG dwarf galaxies found in isolation\footnote{Here we consider as "isolated systems" those LG dwarf galaxies that are unlikely to have had more than one pericenter passage around the large LG spirals, as determined from their free-fall time.} is therefore valuable, as these offer a cleaner view than satellite galaxies into the internal mechanisms that have shaped the evolution of systems at the low end of the galaxy mass function. However, due to their larger distance from us, it is challenging to obtain large spectroscopic samples of individual RGB stars in these objects, and thus the internal kinematics and metallicity properties of their evolved stellar component  have started to be explored only relatively recently \citep[see e.g.][]{Fraternali2009,leaman2013, kirby2014, kacharov2017, taibi2018}. 
 
Pinning down the characteristics of isolated LG dwarf galaxies can also inform models that try to explain the LG so-called "morphology-density relation", i.e., the fact that isolated LG dwarf galaxies mostly contain HI gas ("late-types"), in stark contrast with satellite galaxies of M31 or the MW, which are for the great majority devoid of neutral gas ("early-types").  

Historically, LG late-type dwarfs have been divided into dwarf irregulars (dIrrs) and transition-types (dTs) \footnote{It is beyond the scope of this paper to review the taxonomy of dwarf galaxies, for which we refer the readers to other works in the literature \citep[e.g.][]{tolstoy2009,Ivkovich2019}. Note that there are no Blue Compact Dwarfs in the LG, with the possible exception of IC~10.}. The dTs contain a neutral gas component but no on-going star formation, so that their properties are intermediate to those of dIrrs and early-types such as dwarf spheroidals (dSphs). Because of this, they have been suggested as an evolutionary link between these two types \citep[see e.g.,][]{tolstoy2009}. On the other hand, dTs have also been considered as an extension of dIrrs with low star formation rate \citep{Weisz2011,koleva2013}.
It should be emphasised that the morphological classification is generally based on the dwarfs present-day properties, while it has been shown that dwarfs of the same type may have had a different evolutionary past, as derived from their full star formation history \citep{gallart2015}. This underlines the importance of referring to dwarf galaxies on the basis of their physical properties.
    
This work is part of a series of articles in which we make use of spectroscopic data of individual red giant branch stars to improve the observational picture of the properties of the evolved stellar component of isolated LG dwarf galaxies \citep[][Taibi et al. in prep.]{kacharov2017, taibi2018}. Here we focus on Aquarius (DDO~210; see Table~\ref{table:aqu} for a summary of its main properties). It is located near the edge of the LG, at approximately 1\,Mpc away from the MW and 1.1\,Mpc from M31 \citep[for studies of distance based on RGB stars see][]{vandenbergh1979,lee1999,mcconnachie2005}, and only two galaxies are found within a distance of 500 kpc from it \citep[SagDIG and VV124; ][and references therein]{cole2014}. 
Its free-fall time to the barycenter of the LG is approximately equal to one Hubble time \citep{mcconnachie2006,mcconnachie2012}, which means that it has likely not interacted with M31 or the MW during its lifetime. Moreover, the estimator of the tidal interactions of this galaxy with its closer neighbors is consistent with a system in isolation following the criteria from \citet{Karachentsev2004,Karachentsev2013}. 
    
The classification of this galaxy has varied: it has been referred to both as a dT \citep[e.g.][]{mateo1998, mcconnachie2006} and as a dIrr \citep[][although they recognized its transition properties]{cole2014}. Indeed, it shows a higher gas mass to stellar mass fraction with respect to the other systems classified as dTs, like Pegasus or Phoenix \citep[see e.g.][]{mcconnachie2012}. It has a clear UV surface brightness profile \citep{lee2009} that makes it more akin to systems classified as a dIrr. 
On the other hand, even though it has experienced a very prolonged star formation history, its star formation rate was higher between 6 and 8 Gyrs ago and has declined during the last 2 Gyrs, being almost null currently \citep{cole2014}.

Stars in this galaxy were resolved for the first time by \citet{marconi1990} reaching a magnitude of 23.5 in the V band. The stellar component presents a well-defined position angle and an ellipticity varying with radius, becoming more circular in the outer parts \citep{mcconnachie2006}. The young stars (main-sequence and blue-loop stars), which are the least numerous, present a different surface density profile with respect to the older population (RGB and red clump stars). This indicates that the spatial distribution of star forming regions has varied with time \citep{mcconnachie2006}. As for spectroscopy of the evolved stellar component, there are only two studies that have derived l.o.s. velocities and metallicities of individual stars \citep{kirby2014,kirby2017}.
    
The neutral gas properties are well-determined, showing that the morphologies of HI gas and stars differ \citep[e.g.][]{young2003,begum2004,mcconnachie2006}.
When overlaying the HI contours on optical images of Aquarius, it is clearly seen that they are not coincident \citep{young2003}. This is mainly caused by the position of the young stars in the galaxy, shifted a few arcminutes to the east with respect to the center, and coincident with a small cavity in the HI profile as indicated by \citet{mcconnachie2006}. Although fewer, the young stars are brighter than the older ones, disproportionately impacting the surface brightness profile. This difference in the stellar distribution is probably causing the variation of ellipticity with radius.
Both \citet{young2003} and \citet{begum2004} reported a small velocity gradient in the HI gas and, more recently, this was confirmed by \citet{iorio2017}, who measured the velocity gradient along a P.A. $= 77$ deg. 
    
Here we present results of our study of the chemical and kinematic properties of the stellar component of the Aquarius dwarf, based on VLT/FORS2 MXU spectroscopic observations in the region of the nIR CaT for a sample of 53 individual red giant branch (RGB) stars. The article is structured as follows. 
In Sect.~\ref{sec:reduction} we present the data acquisition and observational details. In Sect.~\ref{sec:velocities} we describe the reduction process and the determination of velocities and metallicities for the whole sample. In Sect.~\ref{sec:kinematics} we apply selection criteria to identify likely member stars and we perform the analysis of the kinematic properties of Aquarius. In Sect.~\ref{sec:metallicity} we analyze the metallicity ([Fe/H]) distribution and explore the possible presence of two different chemo-dynamical stellar populations. In Sect.~\ref{sec:discussion} we discuss the main results and compare with the characteristics of the neutral gas. The summary and conclusions are in Sect.~\ref{sec:conclusions}.

\section{Observations}
\label{sec:reduction}

    \begin{figure*}
    	\centering
    	\includegraphics[width=\textwidth]{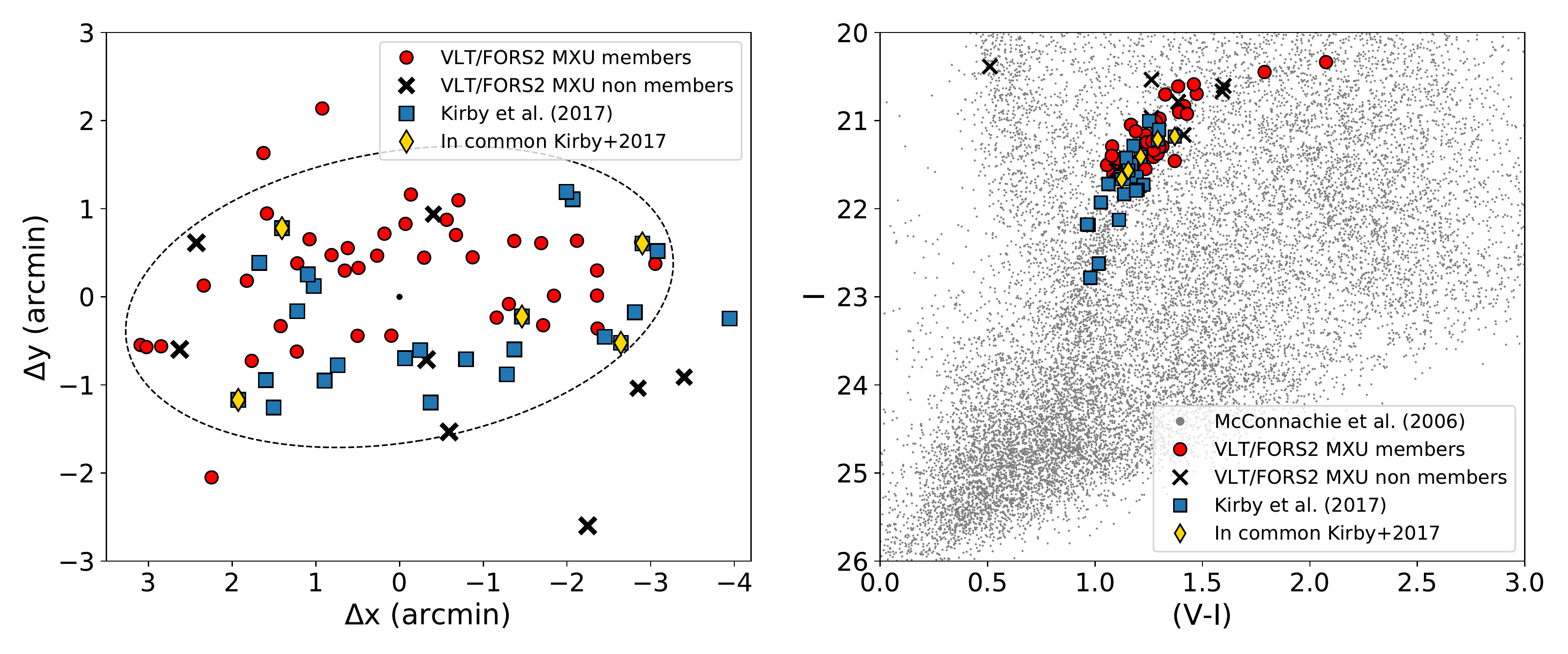}
    	\caption{Spatial distribution (left) of the targets projected onto the tangential plane and color-magnitude diagram of the stars along the line-of-sight to Aquarius (right). The red circles represent the stars observed with VLT/FORS2 MXU and classified as Aquarius's members, blue squares show the sample of stars members from \citet{kirby2017}, yellow diamonds are the stars in common between these two studies, and black crosses correspond to the VLT/FORS2 MXU RGB stars that have been classified as probable non-members of the galaxy. On the left panel, the ellipse has a semi-major axis equal to 3 times the half-light radius of the galaxy, with a position angle of 99$^{\circ}$ and ellipticity of 0.5 (see Table~\ref{table:aqu}). The black dot represents the galactic center. On the right panel, grey points represent the objects classified with high confidence as stars in the Subaru/SuprimeCam photometric data ($34'\times27'$). Magnitudes have been corrected for extinction assuming a uniform Galactic screen and adopting E(B-V) from Table~\ref{table:aqu} along with reddening law $A_{V} = 3.1 \times$ E(B-V).}
        \label{fig:dist}
    \end{figure*}

The data were obtained in service mode between June and September 2013 as part of the ESO Program 091.B-0331 (PI: G. Battaglia) using the FORS2 instrument at UT~1 of the Very Large Telescope (VLT). The targets were selected to have magnitudes and colors consistent with being red giant branch (RGB) stars at the distance of the Aquarius dwarf galaxy. To that aim, we used Subaru/SuprimeCam imaging data in the Johnson-Cousins V- and I- band by \citet{mcconnachie2006} from objects classified as point-sources with high confidence. 
Slits to which we could not assign likely Aquarius RGB stars were allocated to random objects\footnote{This resulted in three of the targets being classified as stellar objects in the I band, but not in the V band.}. To ensure precise slit allocations, we used short pre-imaging exposures obtained with FORS2 within the same program.
    
Figure~\ref{fig:dist} shows the spectroscopic targets' spatial distribution (left) and location onto the Subaru/SuprimeCam color magnitude diagram (right). We obtained 55 spectra for 53 individual objects distributed over two FORS2 MXU masks, each observed with 10 exposures, for a total of 25h. On average the airmass was around 1.1 and the seeing about 0.9\arcsec\ for one of the two masks (Aquarius 0) and 1.1\arcsec\ for the other (Aquarius 1). We refer the reader to the observing log in~\ref{table:log} for more details. 

We adopted the same instrumental set-up and observing strategy as in  \citet{kacharov2017} and \citet{taibi2018} (hereafter T18), where the chemo-dynamical properties of the stellar component of the Phoenix transition type galaxy and the Cetus dwarf spheroidal galaxy have been studied. 
Mask slits were designed to be 1\arcsec\ wide by 8\arcsec\ long (for two slits, 6\arcsec\ lengths were used to avoid overlap with adjacent slits). We used the 1028z+29 holographic grism in conjunction with the OG590+32 order separation filter to cover a wavelength range between $7700-9500$ \AA. This encompassed the region of the near-IR Ca~II triplet (CaT) lines. The spectral dispersion was 0.84 \AA\,pix$^{-1}$ and the resolving power $R=\lambda_{cen}/\Delta \lambda =2560$ at  $\lambda_{cen}=8600$\AA.
Calibration data (biases, arc lamp, dome flat-field frames) and slit acquisition images were acquired as part of the FORS2 standard calibration plan.

    \begin{table}
	    \caption{Parameters adopted for the Aquarius (DDO~210) dwarf galaxy. P.A.$_{*}$ is the position angle, measured from North to East, of the stellar component, while P.A.$_{\rm HI}$ is the P.A. of the kinematic major axis of the HI component; V$_{\rm sys}$ and $\sigma_{*}$ are, respectively, the systemic velocity and velocity dispersion of the stellar component; k and 
	    P.A.$_{\rm rot}$ are the velocity gradient of the stellar component and its position angle. All these are measured for the preferred kinematical model of the stellar component (see Sect.~\ref{sec:rotation}). <[Fe/H]> is the median of the distribution of individual [Fe/H] values, while MAD$_{[Fe/H]}$ is the MAD of the distribution (see Sect.~\ref{sec:metallicity}).}
    		\label{table:aqu}      
		\centering          
		\begin{tabular}{c c c}    
			\hline\hline
			Parameter & Value  & Reference \\ 
			\hline           
			$\alpha_{J2000}$ & $20^h46^m51.8^s$ & (1) \\
			$\delta_{J2000}$  & $-12^{\circ}50'53''$ & (1) \\
			ellipticity & 0.5$\pm$0.1 & (3) \\
			P.A.$_*$ (\degr)& $99\pm1$ & (3) \\
			$R_{h}$(\arcmin) & $1.10\pm0.03$ & (2) \\
			$M_V$ & $-10.6\pm0.3$ &(3) \\
			E(B-V) & 0.045 & (5) \\
			$(m-M)_0$ & $24.95\pm0.10$ & (4) \\
			$D_\odot$ (kpc) & $977\pm45$ & (5) \\
			P.A.$_{\rm HI}$ (\degr)& $77.3\pm15.2$ & (6) \\
			V$_{sys}$ (km\,s$^{-1}$) & $-142.2\pm1.8$ & (7) \\
            $\sigma_{*}$ (km\,s$^{-1}$) & $10.3^{+1.6}_{-1.3}$ & (7) \\
            k (km\,s$^{-1}$\,arcmin$^{-1}$) & $-5.0^{+1.6}_{-1.9}$ & (7) \\
            P.A.$_{\rm rot}$ (\degr) & 139.0$^{+17.4}_{-26.8}$ & (7) \\
            <[Fe/H]> (dex) & -1.59$\pm$0.05 & (7) \\
            MAD$_{[Fe/H]}$ (dex) & 0.20 & (7) \\
			\hline        
		\end{tabular}
		\tablebib{(1) \citet{vandenbergh1959}; (2) \citet{mcconnachie2006};
	    (3) \citet{mcconnachie2012}; (4) \citet{cole2014}; (5) \citet{schlafly2011}; 
	    (6) \citet{iorio2017}; (7) this work.}
	\end{table}

\section{Data reduction process and measurements}
\label{sec:velocities}

We adopted the same procedures described in \citet{kacharov2017} and T18 for the data reduction, as well as  for the determination of line-of-sight (l.o.s.) velocities and metallicities. 
 
The optimally-extracted, background-subtracted 1D spectra for each aperture were corrected for zero-point shifts due to the different date of observation, and possible small slit-centering shifts. Furthermore, the wavelength calibration was refined exploiting the presence of numerous OH telluric emission lines.

The refinement to the wavelength calibration was obtained with the IRAF\footnote{IRAF is the Image Reduction and Analysis Facility distributed by the National Optical Astronomy Observatories (NOAO) for the reduction and analysis of astronomical data. \url{http://iraf.noao.edu/}} \textit{fxcor} task, through cross-correlation between a reference sky spectrum and the OH emission lines visible in the scientific exposures over the wavelength range $8200-9000$\AA. The corrections ranged from $\pm 1.2-16.9$\,km\,s$^{-1}$ with errors $\pm 0.7-2.9$\,km\,s$^{-1}$. 
The slit-centering correction was calculated by comparing the position of the centroid of each slit with respect to the centroid of the stellar flux. The latter is measured in the so-called through-slit images. If the target is not well-centered on the slit, this causes a velocity offset (systematic wavelength shift) for targets that are smaller or comparable
in size to the slit width \citep{Irwin2002}. 
The computed correction ranged from $\pm 0.15-8.5$\,km\,s$^{-1}$ with errors $\pm 0.4-4.2$\,km\,s$^{-1}$. The short slit length of the two $6''$-long slits prevented a proper calculation of the slit-centering shift. Since the value of the slit-centering correction was found to change smoothly as a function of location on the chip, for these two stars we adopted the slit-centering correction of the adjacent targets. The barycentric correction was obtained using the IRAF \textit{rvcorrect} task. All the shifts were applied to the individual spectra using the IRAF task \textit{dopcor}.

\begin{figure}
    	\centering
    	\includegraphics[width=\columnwidth]{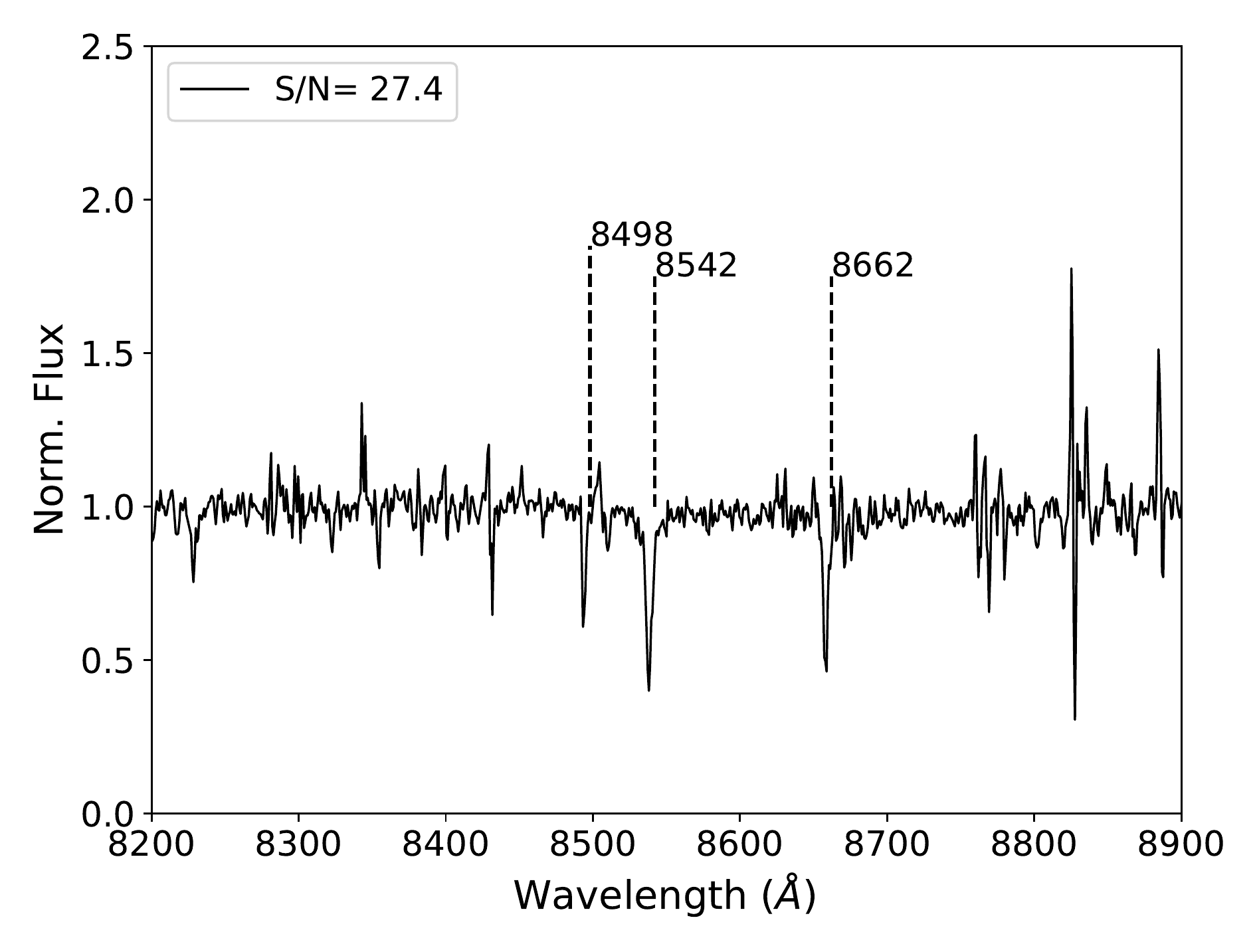}
    	\includegraphics[width=\columnwidth]{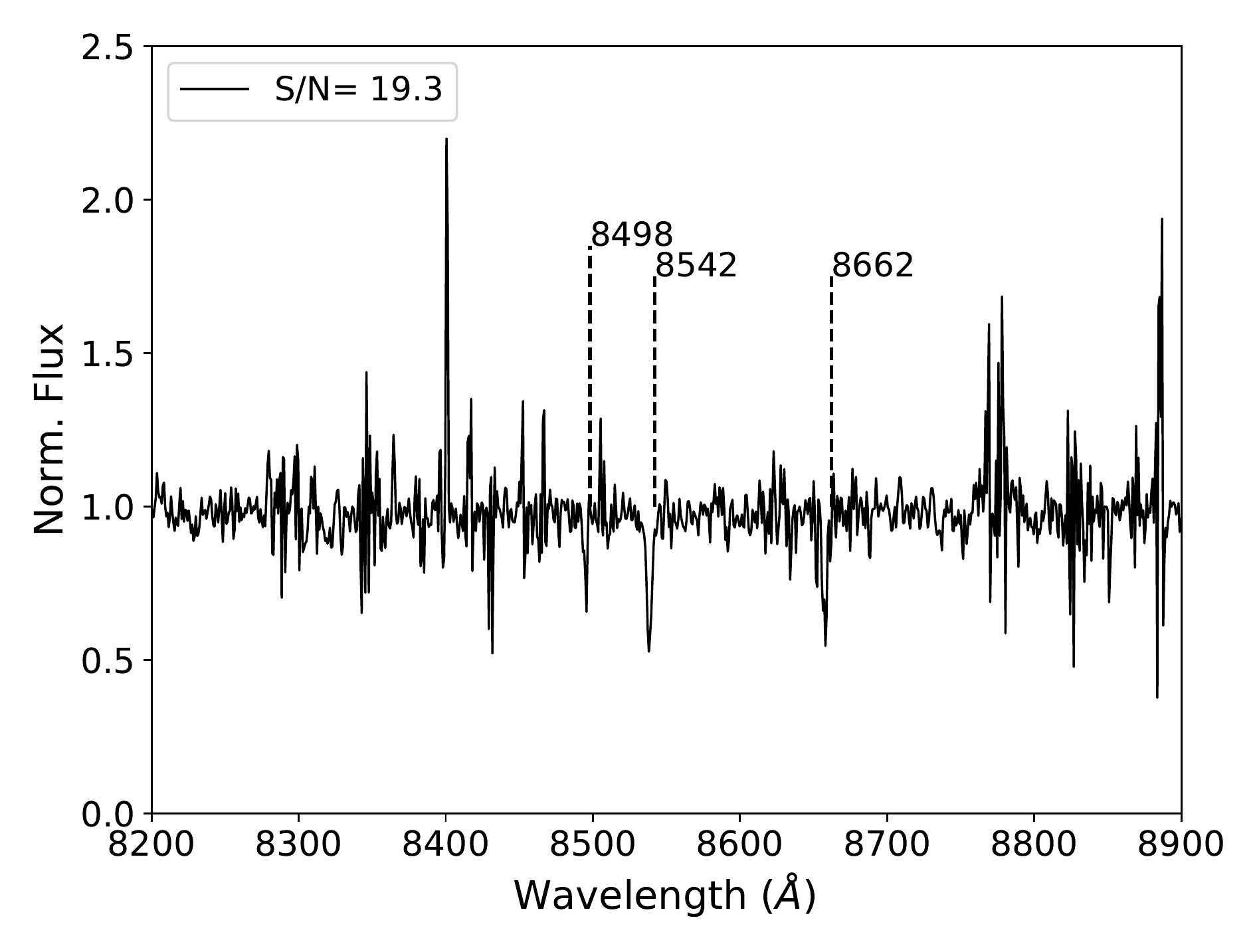}
    	\caption{Example spectra of the stars \textit{aqu0c1star2} (upper panel) and \textit{aqu1c2star3} (lower panel). CaT lines and the S/N for each spectra are indicated.}
        		\label{fig:spectra}
    \end{figure}

The heliocentric velocities, $v_{\rm hel}$, and equivalent widths (EWs) for the CaT lines were measured on the stacked spectra, obtained from a weighted sum of the individual exposures. 
Table~\ref{tabla:stars_aqu} lists the S/N per pixel for each star, measured in the CaT region from the stacked spectra; the median S/N per pixel is 26. Figure~\ref{fig:spectra} shows an example of two stacked spectra. 
    
To calculate the heliocentric line-of-sight velocity of the stars we used again the IRAF task \textit{fxcor} and cross-correlated the stacked spectra with an interpolated Kurucz stellar atmosphere model over the wavelength range between $8400-8750\,\AA$. The model was convolved to have the same dispersion as our data, and its parameters were chosen to represent a low metallicity RGB star: ${\rm log(g)} = 1.0$, T$_{\rm eff}= 4000$\,K, ${\rm [Fe/H]} = -1.5$\,dex.

The CaT EWs were obtained from the continuum normalized stacked spectra, by fitting a Voigt profile to the individual CaT lines, integrating their flux over a window of 15\AA\, and adopting the corresponding error-spectra as the flux uncertainty at each pixel in the fitting process. To obtain estimates of the stars' metallicity ([Fe/H]) we adopted the \citet{starkenburg2010} relation as a function of the (V-V$_{\rm HB}$), linearly combining the EW of the two strongest CaT lines. The errors were calculated by propagation of the EW uncertainties. As a value for V$_{\rm HB}$ we use $25.45\pm0.20$, estimated with photometric data from \citet{cole2014}; we verified that adopting the \citet{starkenburg2010} calibration expressed as a function of the stars absolute visual magnitude leads to the same results. We note that CaT lines have been widely used to estimate the [Fe/H] of RGB stars in a variety of stellar systems, from MW globular and open clusters \citep[see e.g.][]{rutledge1997,carrera2012} to LG dwarf galaxies \citep[see e.g.][]{tolstoy2001,battaglia2008}, and tested and calibrated over a broad range of metallicities and stellar ages \citep[see e.g.][]{battaglia2008,starkenburg2010,carrera2013}. Specifically, the validity of the \citet{starkenburg2010} relation has been tested over the range $-4 \leq {\rm [Fe/H]} \leq -0.5$.

Two stars in our sample have been observed with both masks. The velocity and [Fe/H] for the two measurements were consistent within the errors in both cases. Therefore, their velocities and [Fe/H] have been combined using a weighted mean for the final analysis, leading to a total number of 53 targets.
    
Table~\ref{tabla:stars_aqu} reports the slit information, RA-DEC coordinates, V- and I-band magnitudes, velocity, [Fe/H], S/N ratio, and the membership status according to the criteria applied in Sect.~\ref{sec:kinematics} for each of the targets. The median error in the velocity and [Fe/H] measurements is 4.8\,km\,s$^{-1}$ and 0.13\,dex respectively.

The spectra of four of our targets, labeled as "C" in Table~\ref{tabla:stars_aqu}, contain strong CN bands \citep[see also][]{kirby2017}. For these we closely inspected the results from \textit{fxcor}, which delivered a well-defined cross-correlation peak. 
Based on this and given that \begin{enumerate*}[label=(\roman*)] \item the velocities of these stars are close to the systemic velocity of the galaxy, \item their magnitudes are compatible with belonging to the RGB and \item their metallicities do not stand out from the rest, \end{enumerate*}
we decided to include them in the final sample because they are not biasing significantly our results.

\subsection{Comparison with Kirby et al. (2017)}    

\citet{kirby2017} (hereafter K17) have used Keck/DEIMOS spectroscopic data to measure heliocentric line-of-sight velocities (metallicities) for 25 (23) stars classified as Aquarius members. A search of matches within 2\arcsec\, returns 5 overlapping stars with our sample. 
Table~\ref{table:kirby} lists the heliocentric velocities and [Fe/H] values for these stars in common: the star displaying the largest error in l.o.s. heliocentric velocity according to the determination of Kirby has been measured twice in our sample (once with each mask), yielding measurements compatible within 1$\sigma$ (see previous section) and has been combined in the analysis (\textit{aqu0c1star11}). The velocity derived for this star is compatible within 1$\sigma$ in both studies. However, the measurements for the other 4 stars suggest the presence of a possible zero-point offset ($\sim$10\,km\,s$^{-1}$) between the two studies. For perfectly derived velocity errors, one would expect the distribution of normalized velocity differences, $(v_{\rm FORS2}-v_{\rm DEIMOS})/\sqrt{\delta_{v,\rm FORS2}^2 + \delta_{v, \rm DEIMOS}^2}$, to yield a mean of ($0\pm0.44$) and a standard deviation of ($1\pm0.34$) for a sample of 5. In this case, the mean and standard deviation are 1.75 and 1.8, respectively, supporting the possibility that the bulk of the velocity differences is not due to random errors. We stress however that, due to the small number of overlapping stars, it is difficult to ascertain whether the only source of these differences is a systematic offset.
    
Offsets between radial velocity determinations of the same stars from different studies are not unusual. These happen not only when different spectrographs are used (e.g. see Gregory et al. \citeyear{gregory2019} comparison of FORS2 multi-slit versus FLAMES/GIRAFFE fibres observations of the Tucana dSph) but sometimes even when the same instrument and configuration mode is adopted (e.g. comparison of Keck/DEIMOS $R\sim6500$ observations of the Triangulum~II system, Kirby et al. \citeyear{kirby2017b}). 
However, we have validated our data-reduction and analysis procedure in multiple studies that included sub-samples of repeated exposures for the same stars, which yielded consistent results. Since our methodology has not changed, we are confident that internally our velocity determinations are reliable. 
    
As for the metallicities, the FORS2 and DEIMOS measurements agree with each other within 1-2$\sigma$ for 4 out of 5 stars. Given the different methods used (CaT EWs in one case and spectral synthesis excluding the CaT in the other one), this is very encouraging.
Star \textit{aqu0c1star11}, which is the one that deviates the most in velocity, shows a large deviation in metallicity too. The S/N of the DEIMOS spectra of this star is much lower than the typical S/N of the spectra for the rest of the sample (6\AA$^{-1}$\, versus a mean of $\sim$18\AA$^{-1}$, with a corresponding dispersion of 5.7\AA\,pix$^{-1}$), while its metallicity error is instead similar to that of the rest of the sample (0.16\,dex). It is possible that the error quoted by K17 is underestimated for this star. In support of this statement, we note that there is another star in the K17 sample with a spectrum of S/N = 6\AA$^{-1}$\, and that has instead an error in [Fe/H] of 0.45\,dex.
    
\section{Kinematic analysis} \label{sec:kinematics}
    
\subsection{Membership}

In order to determine the kinematic and metallicity properties of the galaxy, we had to apply selection criteria to discard foreground contaminant stars along the line of sight to Aquarius. 
    
\begin{figure}
    		\centering
    		\includegraphics[width=\columnwidth]{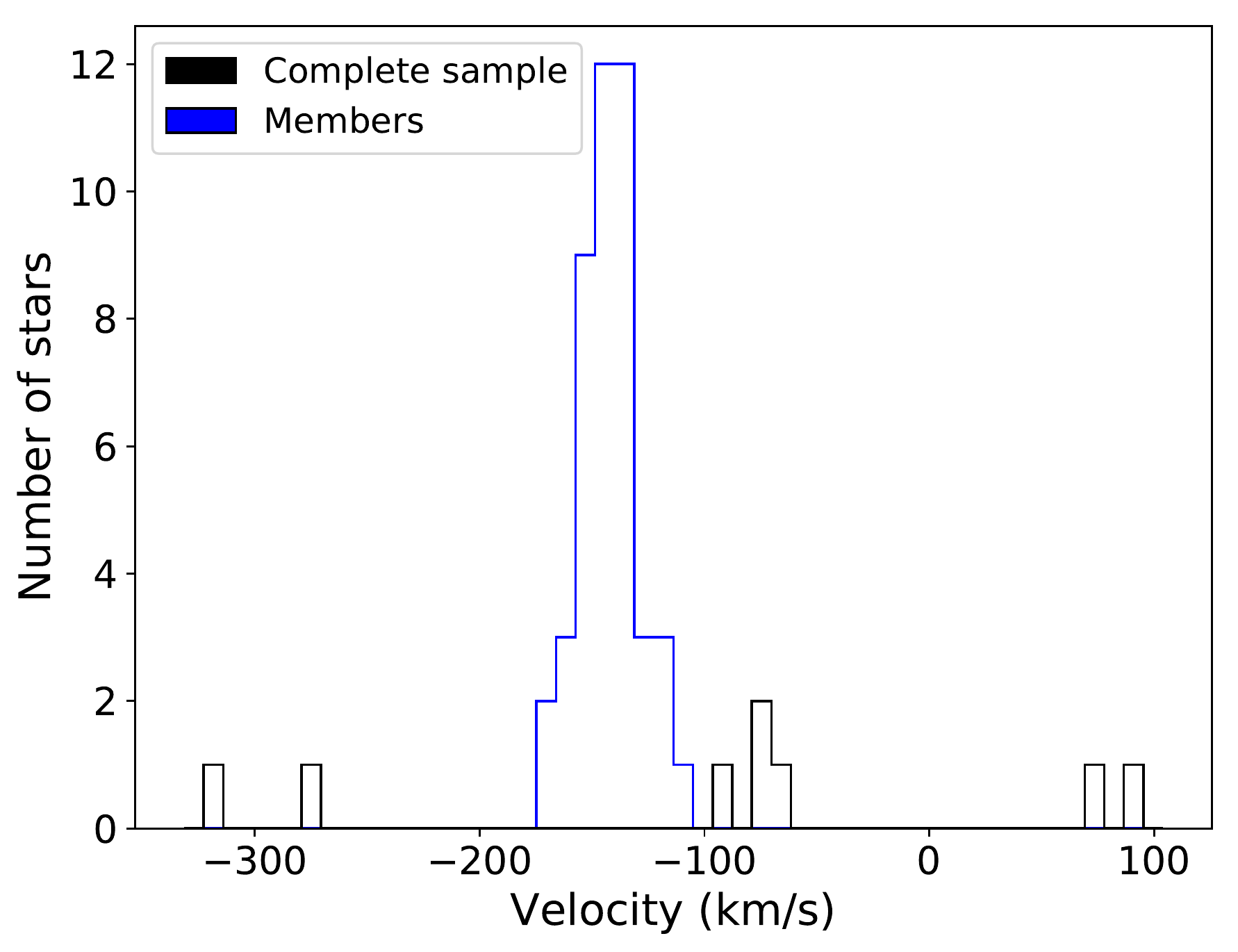}
    		\caption{Histogram of the heliocentric velocities of all the targets (black) and the 45 targets that were finally considered as members of the galaxy (blue).}
    		\label{fig:finalhistvel}
   \end{figure}

We excluded the stars in two steps. First, we eliminated one star whose magnitude and color is not compatible with RGB stars at the distance of Aquarius.
Then, we performed a selection on the basis of the heliocentric velocity of the stars.
We adopted $\mid v_{\rm hel} - \bar{v}_{\rm hel} \mid \leq 3$\,MAD($v_{\rm hel}$) as a simple approach to exclude the targets\footnote{The median absolute deviation is defined here as an estimator of the standard deviation: MAD$(X)= 1.48 \times median(\left | X - median(X) \right |)$}. 
The estimation of these parameters was an iterative process, fitting the data until convergence, and including the possible presence of rotation (see Sect.~\ref{sec:rotation}) at the same time. The data-set was reduced from 52 to 46 targets. 
We double checked this kinematic selection by applying a Bayesian analysis (see also Sect.~\ref{sec:rotation}) that solved iteratively the systemic velocity of the system $v_{\rm sys}$, the velocity dispersion $\sigma_v$ and the best-fitting model for the internal kinematics (dispersion-only or dispersion + rotation). We used this information to calculate the expected bulk velocity at the position of each given star ($v_{\rm bulk}$) and retain those targets that fulfilled the condition $\mid v_{\rm hel} - v_{\rm bulk} \mid \leq 3$\,MAD($v_{\rm hel} - v_{\rm bulk}$). This process excluded one more star from the sample, giving us a final number of 45 members.
A histogram of the velocities for all targets and members is shown in Fig.~\ref{fig:finalhistvel}. 
    
   \begin{figure}
	\centering
	\includegraphics[width=\columnwidth]{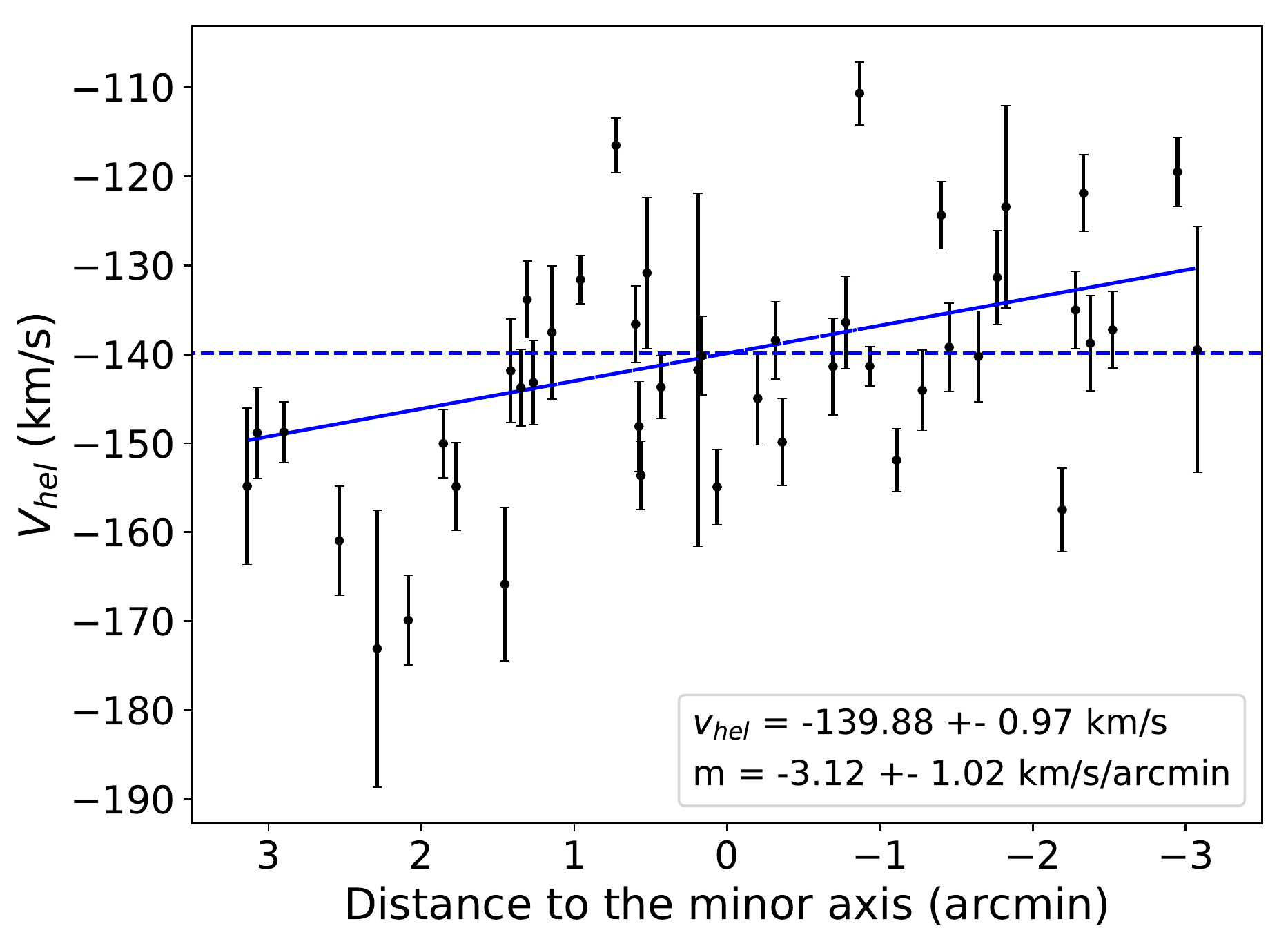}
    \caption{Velocity distribution of the probable Aquarius members with respect to their distance 
            to the minor axis. The blue solid line shows the best weighted linear fit to the 
            stars' velocities and the dashed line the systemic velocity derived from the fit.}
    \label{fig:rotation}
    \end{figure}
   
\subsection{Internal kinematic properties}
\label{sec:rotation}

\citet{wheeler2017} carried out a systematic analysis of the rotational support of LG dwarf galaxies, based on line-of-sight velocities from literature studies. One of the main aims of the authors was to understand whether the rotational support of the stellar component of LG late-type dwarf galaxies is significantly different from that of dSph satellites of the MW and M31, as would be predicted by the "tidal stirring model" put forward to explain the morphology-density relation of the LG \citep{Mayer2001,Mayer2006,kazantzidis2011}. For a full discussion of the importance of determining the internal kinematic properties of different classes of dwarf galaxies, we refer the reader to, for example, \citet{wheeler2017} and \citet{Ivkovich2019}. Here we explore such properties for our VLT/FORS2 sample of Aquarius' stars.

Figure~\ref{fig:rotation} offers a first look into this aspect by displaying the heliocentric line-of-sight velocities of Aquarius member stars as a function of the distance from the projected optical minor axis of the galaxy (see Table~\ref{table:aqu}): a velocity gradient is clearly visible, with a weighted linear fit yielding a slope of $-3.1\pm1.0$\,km\,s$^{-1}$\,arcmin$^{-1}$.
The systemic velocity and velocity dispersion obtained are $v_{\rm sys}=-139.9 \pm 1.0$ \,km\,s$^{-1}$ and $\sigma_{\rm v}$=11.0\,km\,s$^{-1}$, respectively.
A weighted linear fit to the $v_{\rm hel}$ along the minor axis instead is consistent with no velocity gradient.
Therefore there are indications of a mild amount of rotation in this system. We note that Aquarius is an isolated galaxy with a small angular extent on the sky, so that it is highly unlikely that the detected velocity gradient is due to effects such as tidal disturbances or projection effects of the 3D motion of the galaxy across the line-of-sight. 

We perform a Bayesian analysis in order to search for the presence of velocity gradients in Aquarius without fixing a preferred axis a priori (for more details on the methodology see T18). The corresponding results are presented in Table~\ref{table:multinest}. We compare three different models:
a dispersion-only model and a model including both random motions and rotation, expressed either as linear rotation or constant (flat) rotation velocity as a function of radius. The free parameters are: the systemic velocity and velocity dispersion in the three models (v$_{\rm hel}$ and $\sigma_{\rm v}$, respectively); the position angle ($\theta$) of the kinematic major axis for the two models with rotation; and, in addition, the slope of the velocity gradient for the linear model ($k$) and the value of the rotational velocity for the flat model ($v_c$).
The results of the three models can be compared in terms of the Bayes factor, i.e. the ratio of the Bayesian evidence of a given model against the other (ln$B_{\rm lin,flat}$ and ln$B_{\rm rot,disp}$). On the Jeffrey scale, when the natural logarithm of the Bayes factor is (0-1),(1-2.5),(2.5-5),(5+) it can be interpreted as inconclusive, weak, moderate and strong evidence, respectively.

We find that the linear rotation model is weakly favoured both with respect to the constant rotation model (ln$B_{\rm lin,flat}$=1.7) and with respect to the dispersion-only model (ln$B_{\rm rot,disp}$=1.6). The systemic velocity, dispersion and slope of the velocity gradient derived with this approach are, respectively, $v_{\rm sys} = -142.2^{+1.8}_{-1.8}$ \,km\,s$^{-1}$, $\sigma_{\rm v}$ = 10.3$^{+1.6}_{-1.3}$ \,km\,s$^{-1}$ and $k = -5.0^{+1.6}_{-1.9}$ \,km\,s$^{-1}$\,arcmin$^{-1}$ ($-17.6^{+5.6}_{-6.7}$ \,km\,s$^{-1}$\,kpc$^{-1}$), consistent within 1-2$\sigma$ from the determinations obtained with a weighted linear fit to the velocities along the major axis. The best-fitting position angle of the kinematic major axis is $\theta \sim139_{-27}^{+17}$\,degrees, shifted with respect to that of the projected major-axis of the galaxy, although consistent with it at the 1.5-$\sigma$ level. In our definition, a negative velocity gradient with $\theta$ between $0^{\circ}$ and $180^{\circ}$ implies a receding velocity on the West side (and would be equivalent to a positive gradient with $\theta = \theta + 180^{\circ}$).

\citet{iorio2017} find a weak velocity gradient in the HI gas: they measure a rotational velocity of $\sim$5\,km\,s$^{-1}$ out to a radius of 1.5\arcmin\, (using the distance they assumed to convert from kpc to arcmin), similar to the rotational velocity we would obtain at approximately the same radius\footnote{As a side note, the authors quote the maximum circular velocity within the observed radial range, as the data for Aquarius was not reaching parts of the galaxy far enough out to include the flat part of the rotation curve. The value for the circular velocity, after the much dominating asymmetric-drift correction, is V$_{0} = 16.4\pm9.5$\,km\,s$^{-1}$.}. 
However, the kinematic major-axis of the HI gas has a P.A. of $77.3\pm15.2$ degrees (receding velocities on the East side), which is misaligned with the kinematic P.A. of the stellar component here examined. 
In Sect.~\ref{mock} we show that this misalignment is unlikely to be a consequence of the characteristics of the FORS2 data-set, in terms of number statistics, spatial coverage and velocity uncertainties, and in Sect.~\ref{sec:discussion} we discuss the possible origins of this feature, placing it in the context of the complexities of Aquarius's structure and HI properties.

\subsection{Comparison with other works}

We have also applied our Bayesian analysis of the Aquarius internal kinematics to the smaller sample of 25 member stars by K17. This sample consists for the great part of re-observations of stars in \citet{kirby2014} (the latter work had 27 members, 24 of which are in common with K17).
    
For the K17 data-set, the comparison between the two rotational models does not clearly favour one over the other (ln$B_{\rm lin,flat}=0.13$), and the comparison between the (only slightly) favoured linear rotation model against the dispersion-only one is ln$B_{\rm rot,disp}=0.28$. Therefore, the presence of rotation cannot be proven conclusively from the K17 sample (see also Sect.\ref{mock}).    
The lack of constraining power of the DEIMOS sample in terms of rotation signal is also consistent with the analysis by K17, who only placed a 95\% confidence limit of a constant rotational velocity to be $<$9\,km\,s$^{-1}$.

The presence of rotation in Aquarius has also been studied in \citet{wheeler2017} using the \citet{kirby2014} data-set. The authors applied a similar Bayesian statistical analysis to ours, comparing a dispersion model, a constant rotation model and a rotational model considering a radially varying pseudo-isothermal sphere.
They found ln$B_{\rm lin,flat}=-1.00$ and ln$B_{\rm rot,disp}=0.62$. The (only weakly) favoured model is the flat rotational model, although the Bayesian evidence on the presence of rotation is inconclusive. 
    
We have also applied our method to the sample of \citet{kirby2014} and found similar results to \citet{wheeler2017} (ln$B_{\rm lin,flat}=-1.74$ and ln$B_{\rm rot,disp}=0.97$), despite the difference in one of the rotational models. We also recover closely the values of the best-fitting parameters as in \citet{wheeler2017}. We note that the K17 data-set gives the same results as the \citet{kirby2014} sample, apart from a shift in systemic velocity.

    \begin{table*}
      \caption{VLT/FORS2 and Keck/DEIMOS l.o.s. velocities and [Fe/H] for the stars in common between this work and \citet{kirby2017}.}
   		\label{table:kirby}
		\centering          
		\begin{tabular}{c c c c c}    
		   \hline\hline
		   Star & v$_{FORS2}\pm\delta$v & v$_{DEIMOS}\pm\delta$v & [Fe/H]$_{FORS2}\pm\delta$[Fe/H] & [Fe/H]$_{DEIMOS}\pm\delta$[Fe/H] \\ 
				 & [\,km\,s$^{-1}$] & [\,km\,s$^{-1}$] &  [\,dex]   &   [\,dex] \\
		   \hline      
		   aqu0c1star11 & -177.0$\pm$9.0 & -147.4$\pm$28.7 & -1.87$\pm$0.12 & -0.91$\pm$0.16 \\
		   aqu0c2star3 & -119.5$\pm$3.9 & -141.7$\pm$2.9 & -1.63$\pm$0.15 & -1.56$\pm$0.12 \\
		   aqu0c2star7 & -124.4$\pm$3.8 & -136.7$\pm$2.4 & -1.32$\pm$0.12 & -1.46$\pm$0.11 \\
		   aqu1c1star10 & -143.2$\pm$4.8 & -148.9$\pm$2.3 & -1.70$\pm$0.13 & -1.42$\pm$0.11 \\
		   aqu1c2star3 & -137.2$\pm$4.3 & -143.2$\pm$3.2 & -1.66$\pm$0.20 & -1.76$\pm$0.15 \\
		   \hline        
		\end{tabular}
	\end{table*}     
    
\subsection{Mock tests}    
\label{mock}

We have performed a series of tests on mock catalogues in order to understand what type of rotational properties can be detected, given the characteristics of the data-sets in hand. To this aim, we have produced sets of mock l.o.s. velocities at the same position as the spectroscopically observed stars, which we have analyzed as for the actual data. These were extracted from Gaussian distributions centered on a $v_{\rm rot, mock}$ and with standard deviation equal to the l.o.s. velocity error corresponding to that given star. The $v_{\rm rot, mock}$ at a (x, y) position are obtained from the given rotational model under consideration (linear or flat), having $v_{\rm rot, mock}$/$\sigma_{\rm mock} = n =  1.5, 1.0, 0.75, 0.5, 0.25, 0$ at twice the half-light radius (see Table~\ref{sec:intro}); the velocity dispersion $\sigma_{\rm mock}$ was fixed to 10\,km\,s$^{-1}$ for all cases. 

The simulated linear rotation models correspond to velocity gradients $k = 6.8, 4.5, 3.4, 2.3, 1.1$ and 0 \,km\,s$^{-1}$\,arcmin$^{-1}$, while the constant rotation models had rotational velocities $v_{\rm c} = 15, 10, 7.5, 5.0, 2.5$ and 0 \,km\,s$^{-1}$.
All the cases have been simulated for three different position angles which correspond to the P.A. of the projected semi-major axis of the stars (99$^\circ$), then adding 45 and 90 degrees (semi-minor axis). Each case was simulated N = 1000 times.
	
\begin{figure*}
    	\centering
    	\includegraphics[width=\columnwidth]{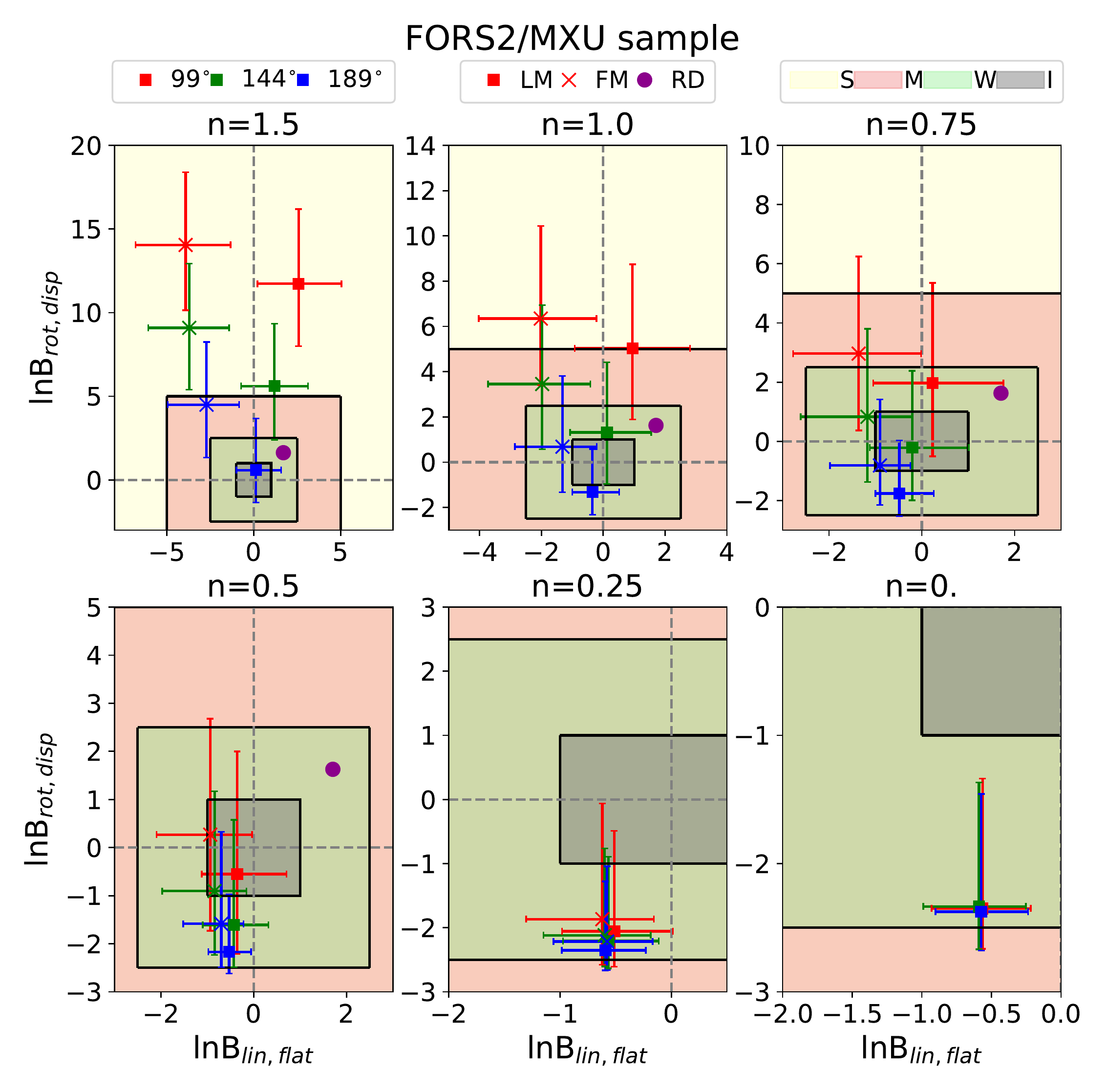}
    	\includegraphics[width=\columnwidth]{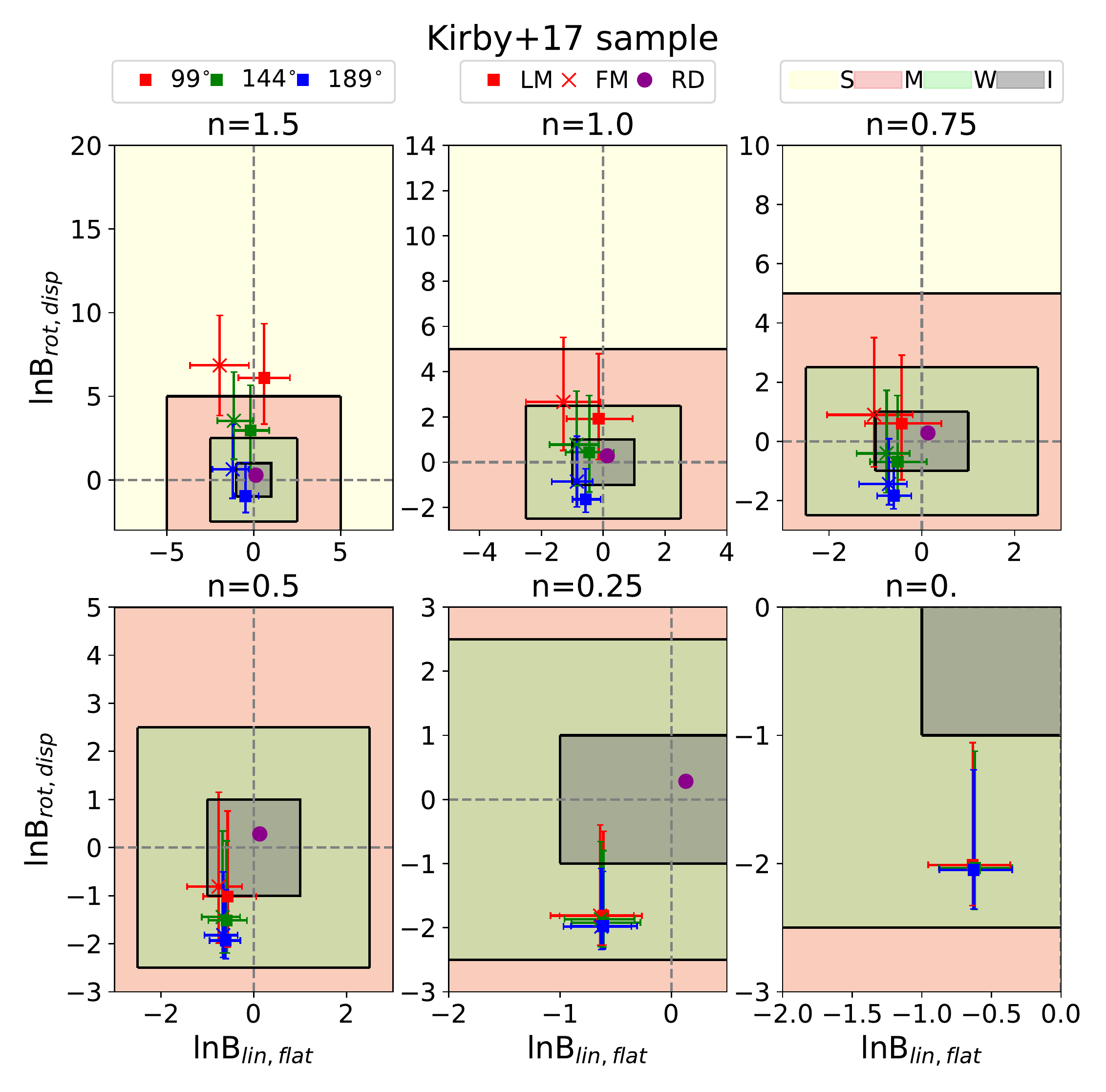}
    	\caption{Bayes factor of the mock tests performed on the FORS2/MXU (left) and K17 (right) samples. Each panel represents the results for a different value of $v_{\rm rot, mock}$/$\sigma$, as given in the panel title. Squares and crosses refer to the linear and flat rotational model (LM, FM), respectively. The color of the markers (red, green and blue) represents a different simulated position angle (99, 144 and 189 degrees, respectively). The purple circle indicates the Bayes factor derived for the real data-sets (we determine the one referring to the K17 sample in Sect.~\ref{sec:kinematics}). Black solid lines discriminate between the strengths of the evidences for each case (yellow: strong; pink: moderate; green: weak; grey: inconclusive).}
        		\label{fig:evidences}
    \end{figure*}

The experiments have been run both for the FORS2 MXU and the K17 samples; the Bayes factors for each $n$ value, position angle and model are shown in Fig.~\ref{fig:evidences}.
	
Results from the tests indicate that rotation can be detected with at least a weak positive evidence when along the optical major axis of the galaxy for $n \ge 0.75$ for the FORS2 data-set and $n \ge 1$ for the K17 sample. Both data-sets have a similar sensitivity to the direction of rotation and model type in terms of ranking, with the FORS2 sample yielding larger evidences for a given model, and therefore a better capability to detect rotation at a given $n$. The data-sets under consideration are not sensitive to the presence of small levels of rotation ($n \le 0.5$ for FORS2, $n \le 0.75$ for K17), since these return either inconclusive evidence or even a weak evidence disfavouring the presence of rotation. In the dispersion-only case ($n=0$) both samples favour this model versus the two models that include rotation, showing that there is no bias towards the rotational models.
	
In terms of strength of the evidence (see Fig.~\ref{fig:evidences}), the value derived for Aquarius stars observed with FORS2 appears compatible with a linear gradient, but not with a flat rotation model. This result could be a consequence of the fact that dwarf galaxies typically have slowly rising rotation curves, which do not always reach their flat part at the last measured point, as it can be appreciated from the kinematics of the HI component \citep[e.g.][]{oh2015, iorio2017}. 
	
The underlying rotation could have rotational support $n=1.5$ along the minor axis or intermediate P.A., $n=1$ along the intermediate axis or the major axis, or $n=0.75$ along the major axis. For the K17 sample, the evidences in favour of one or the other rotational model are not so different from each other, and therefore a distinction does not appear possible. However, in terms of compatibility between observed Bayes factor and levels of $n$ and direction of the gradient that might induce it, the outcome is similar to that of the FORS2 sample.
	
\begin{figure}
    	\centering
    	\includegraphics[width=\columnwidth]{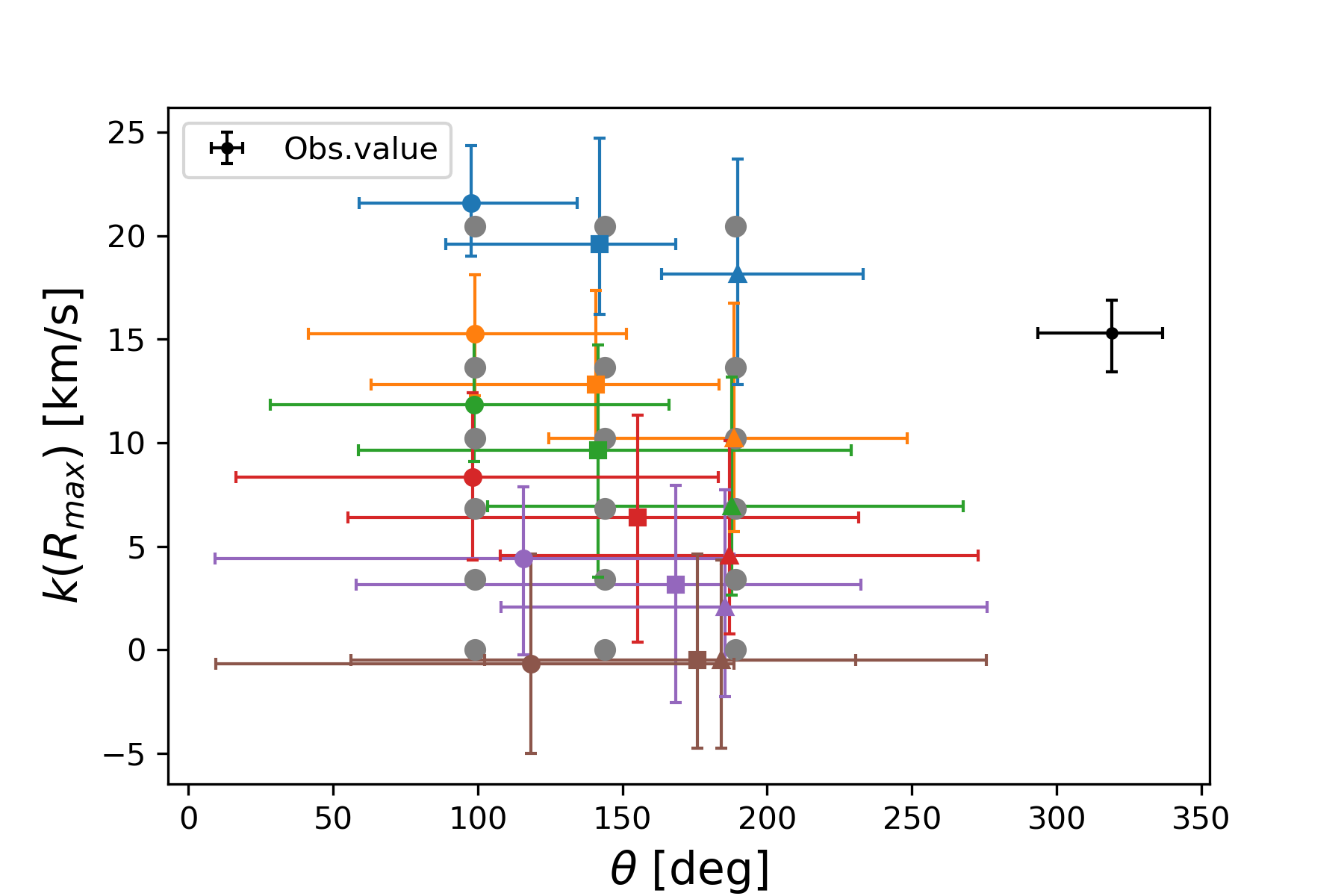}
    	\caption{Rotation velocity at $R_{\rm max}=3'$ vs. kinematic position angle recovered from the tests on mock catalogues reproducing the characteristics of the FORS2 data-set and simulating a linear rotation model. Grey circles represent input values at $n=1.5,1.0,0.75,0.5,0.25,0.$; colored circles, squares and triangles represent recovered values at different simulated kinematic P.A. respectively; colors from blue to brown represent decreasing values of \textit{n}, while the error bars are the 99\% confidence interval (C.I.); in black the observed value from our data-set with error bars at 68\% C.I.}
        		\label{fig:mock_targets}
    \end{figure}

From our tests on the mock catalogues, we also extract the information on how well the kinematic major axis P.A. and the rotational velocity at a given radius are recovered for the various $n$ (see Fig.~\ref{fig:mock_targets}). The initial values are always retrieved within the 99\% confidence interval (C.I.), with a small bias towards underestimating the rotation when the kinematic major axis is along the optical minor axis. Also in terms of amplitude of the rotational velocity, the favourite models would be those with $n\ge 0.75$. 
	
It should be noted that the P.A. of the kinematic major axis for the observed RGB stars is beyond the 99\% C.I. of that retrieved for the models with kinematics similar to that exhibited by the HI gas, indicating that the misalignment/counter-rotation of the stellar component and HI gas is not due to number statistics, spatial coverage and measurement errors of our FORS2 data-set.
    
\section{Metallicity properties}
\label{sec:metallicity}

\begin{figure}
	\centering
	\includegraphics[width=\columnwidth]{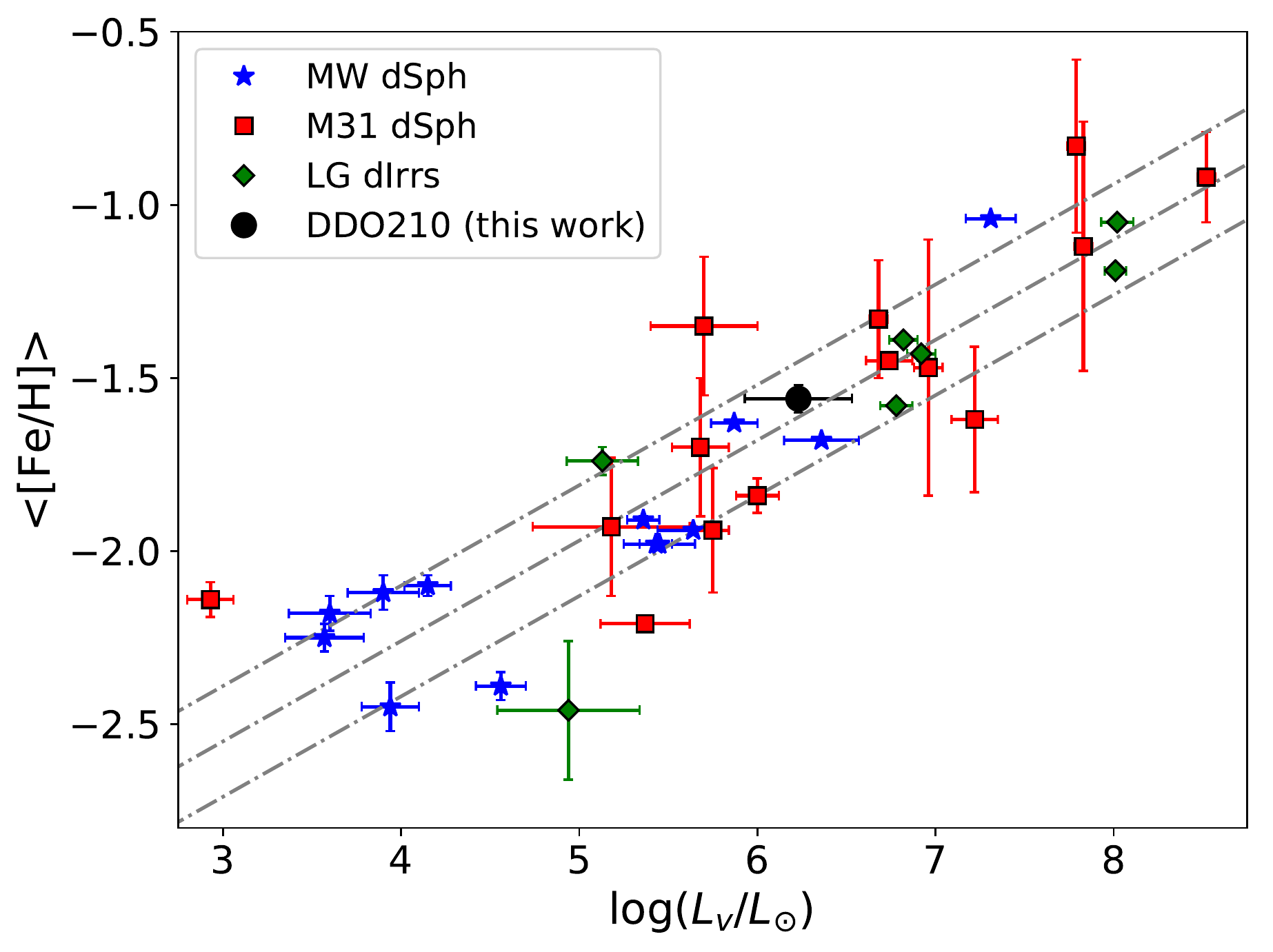}
    \caption{Luminosity-stellar [Fe/H] relation for Local Group dwarf galaxies. 
            Blue stars represent the MW-dSph satellites; red squares are M31-dSph satellites; green diamonds are dIrrs in the LG; the black circle is the position of Aquarius based on the [Fe/H] derived in this work; grey lines are the least-squares linear fit for the dIrrs and MW dSphs and the 0.16 scatter limits. All the values apart from Aquarius were taken from \citet{kirby2013}.}
    \label{fig:lfehrelation}
\end{figure}

The [Fe/H] values obtained for all FORS2 targets are listed in Table~\ref{tabla:stars_aqu}\footnote{We note that these are meaningful only if the star is a giant at the distance of Aquarius, and are not applicable to non-members.}.
We derive the median of the metallicity distribution of all members of Aquarius to be [Fe/H] $= -1.59 \pm 0.05$\,dex. This is compatible with the average value of $<$[Fe/H]$> = -1.50\pm0.06$\,dex by K17 and places Aquarius straight onto the luminosity-metallicity relation for Local Group dwarf galaxies, as can be seen in Fig.~\ref{fig:lfehrelation}. The spread measured as the MAD is 0.20\,dex, while the intrinsic dispersion, taking into account the measurement errors, assuming a Gaussian form for the metallicity distribution function (MDF) is 0.25\,dex. This value is at the low end, but still compatible, with the [Fe/H] dispersion of other Local Group dwarf galaxies \citep[e.g.][]{leaman2013}.

\begin{figure*}
	\centering
	\includegraphics[width=\textwidth]{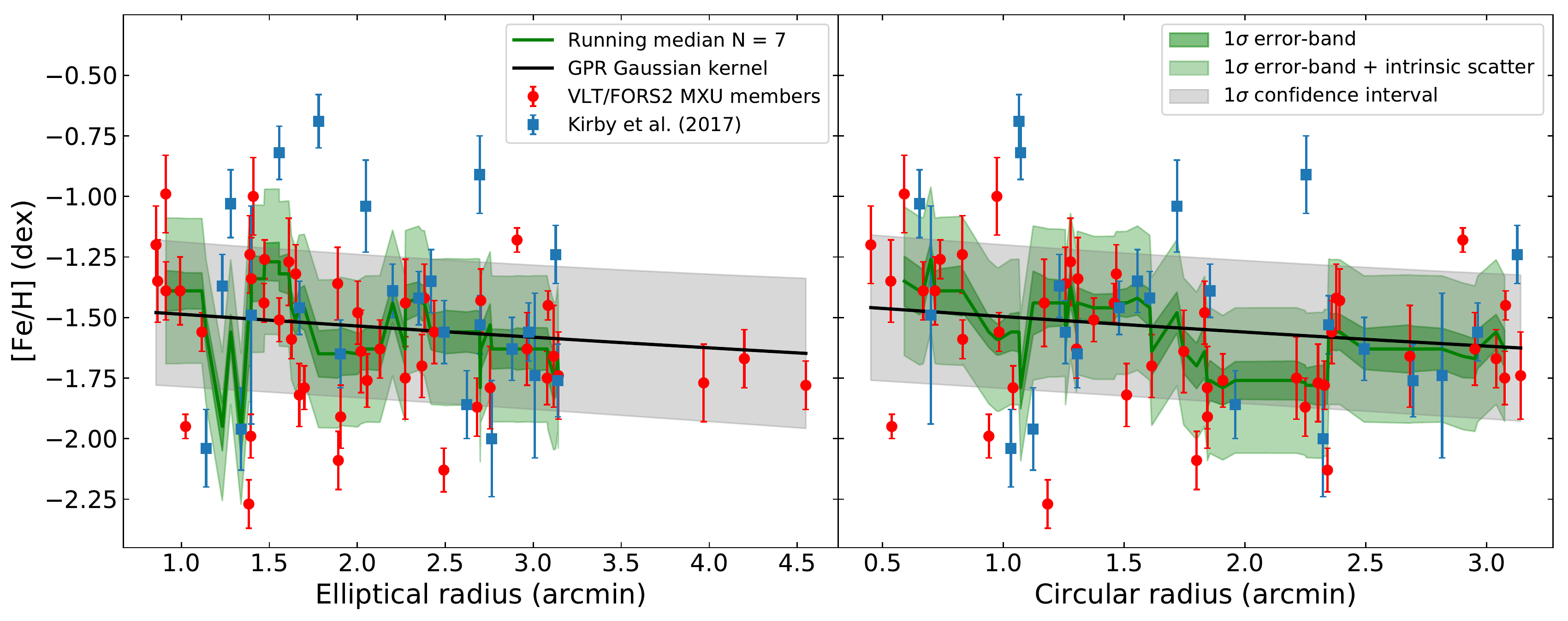}
    \caption{[Fe/H] as a function of the elliptical radius (left) and circular radius (right) for Aquarius member stars observed with FORS2 (red circles) and with DEIMOS \citep{kirby2017} (blue squares). The error-bars show the uncertainties on the [Fe/H] measurements for the individual stars.
    The green solid line represents a running median boxcar with a kernel size of 7 points; the green band shows the 1$\sigma$ error for the running median boxcar, taken as the standard deviation of 1000 Monte Carlo realizations of the running median itself, where the metallicities are extracted from Gaussians centered on the measured [Fe/H] of each star and dispersion given by the measurement uncertainty; the light green band is the same error-band with the MDF intrinsic scatter added in quadrature.
    The black solid line represents the result of a Gaussian process regression analysis using a Gaussian kernel and taking into account an intrinsic scatter; the grey band indicates the corresponding 1$\sigma$ confidence interval.}
    \label{fig:gradient}
\end{figure*}

Figure~\ref{fig:gradient} shows the variation of [Fe/H] as a function of the elliptical (left) and circular (right) radius\footnote{We call "elliptical radius" the semi-major axis of the ellipse that passes through the ($x$, $y$) location of a given star and that has center, ellipticity and P.A. as in Tab.~\ref{table:aqu}; the "circular radius" is instead simply $(x^2 + y^2)^{1/2}$. Here {\it x} and {\it y} are the star's projected celestial coordinates.}. In this figure we include also the points from K17 since we have verified that their metallicity distribution function compares well with that from our FORS2 sample, as in general do the individual [Fe/H] measurements for the stars in common. 
    
It has been shown in the literature that a linear fit does not always fully capture the trend of spatial variations in the metallicity properties of Local Group dwarf galaxies: e.g. in some systems, a decline in the mean metallicity properties is followed by a flattening in the outer parts (see e.g. Sextans, \citet{battaglia2011}; Cetus, T18). Therefore we adopt more flexible ways to determine the general trend as a function of radius, i.e. a running median and a  Gaussian-process (GP) regression. In order to take into account the effect of measurement errors (as well as of the intrinsic scatter), we obtain 1000 MonteCarlo realizations of the individual metallicities, extracting them from Gaussians centered at the observed values and with dispersion given by the metallicity errors (added in quadrature to the intrinsic scatter in the MDF). In the case of the GP regression, we use a Gaussian kernel together with a noise component to take into account the intrinsic scatter of the data. This latter method has the advantage to not depend on a fixed boxcar, like the running median, and the output has a probabilistic meaning. In both cases, the presence of a very mild negative metallicity gradient can be appreciated. This is also seen when considering the run of the metallicity as a function of circular radius (not corrected by the ellipticity). There are hints of a flattening of the slope at large radii, however the current sample size does not allow to place this on a firm ground. We have verified that considering only the FORS2 sample would lead to a very similar trend from the running median or the GP analysis. 

Since the GP analysis returns a trend similar to a very simple linear relation, we perform a simple Bayesian linear regression, including an intrinsic scatter term, to estimate the significance of the metallicity gradient.
The resulting cumulative posterior distribution of the slope of the metallicity gradient indicates that the possibility of no-gradient is within 96\% ($\sim1.75 \sigma$) and 94\% ($\sim1.55 \sigma$) of the distribution, when considering the elliptical radius and the circular radius, respectively. So, while there are indications of a gradient, with this data-set we cannot exclude a flat trend within 2-sigma in both cases. 
    
From an observational point of view, should the presence of a metallicity gradient in an isolated dwarf galaxy such as Aquarius be strengthened in the future with larger data-sets, it would lend further support to the hypothesis that negative metallicity gradients in Local Group dwarf galaxies are not to be ascribed to interactions with the large Local Group galaxies \citep[see e.g. the case of VV~124 and Phoenix,][]{kirby2012, kacharov2017}. Factors such as the dwarf galaxy's star formation history, gravitational potential and rotational versus dispersion support, as well as specific accretion events, are indeed expected to contribute in producing metallicity gradients \citep[e.g.][]{marcolini2008, schroyen2013, benitez-llambay2016, revaz2018}.  As for satellite galaxies, it is possible that effects such as tidal and ram-pressure stripping of the gaseous component could modify the strength of such gradients, exacerbating them depending on the infall time onto the host halo versus the time when star formation ceased, or on the contrary blurring them in the case of strong tidal interactions \citep[][]{sales2010}.   

We defer to a future study the analysis of the possible correlations between rotational support and spatial variations of the metallicity properties of Local Group dwarf galaxies (Taibi et al. in prep.), along the lines of the work by \citet{leaman2013}.
        
The possible presence of a metallicity gradient has led us to look for sub-populations with different chemo-kinematic properties. We divided our data-set into a metal-rich (MR) and a metal-poor (MP) sample based on the median [Fe/H] value of the entire set (22 and 23 stars, respectively). We performed then a Bayesian maximum likelihood analysis (as done in Sect.~\ref{sec:rotation}) on both samples; the resulting parameters and evidences are reported in Table~\ref{table:multinest}. We can see, independently from the fitted kinematic model, that the velocity dispersion values for the two samples are at $\sim$2$\sigma$ from each other. We also note that the evidences of rotation have decreased, due to the low number of targets in each set.
This tentative result of a spatially concentrated metal-rich population with a lower velocity dispersion compared to a spatially extended metal-poor one with a higher dispersion value, adds to what is already found in several other dwarf galaxies of the LG \citep[e.g.][]{tolstoy2004,battaglia2006,battaglia2008,amorisco2012,breddels2014,taibi2018}.
However in our case we would benefit from a larger sample in order to place stronger constraints on the velocity dispersion of the MR and MP stars.

\begin{figure*}
    	\centering
    	\includegraphics[width=\textwidth]{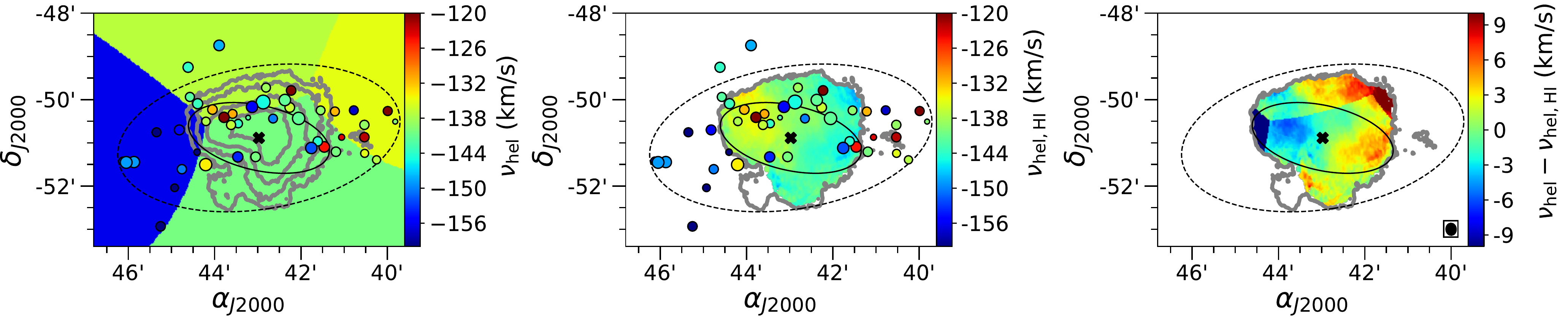}
    	\caption{Comparison between the HI \citep{iorio2017} and stellar (this work) velocity field. Left-hand panel: Voronoi-binned stellar velocity field (SNR$\sim3$, see text for details); the gray contours show the HI surface density at 0.6, 1.2, 1.8, 2.4   M$_{\odot}$/pc$^2$, the lowest contour is at 3$\sigma$ above the noise (also shown in the middle and right-hand panels). Middle panel: HI velocity field. Right-hand panel: Pixel-to-pixel difference between the stellar and HI velocity field; the small ellipse in the bottom right corner shows the beam of the HI observations. In the first two panels the circles represent the stars in our sample: the color indicates their $v_{\rm hel}$ and their size is proportional to $v_{\rm hel}/\delta\,v_{\rm hel}$. In each panel the dashed ellipse is the same as Fig.~\ref{fig:comp_gas}, while the solid ellipse shows the P.A. and the inclination obtained through the analysis of the HI kinematics disc by \citet{iorio2017}. The cross indicates the galactic center (see Table~\ref{table:aqu}).}
          \label{fig:comp_gas}
    \end{figure*}

\begin{table*}
	\caption{Parameters and evidences resulting from the application of the Bayesian analysis to the whole FORS2/MXU sample of members, as well as divided in metal-rich (MR) and metal-poor (MP) samples. We note that the systemic velocity of the HI gas is $-140$\,km\,s$^{-1}$ \citep{iorio2017}, perfectly compatible with our values.}
	\label{table:multinest}
	\centering          
	\begin{tabular}{c c c c c c c|l}    
		\hline\hline
		Sample & Method & $v_{\rm hel}$ & $\sigma_{\rm v}$ & \textit{k} & $v_{\rm c}$ & $\theta$ & Bayes factor \\ 
			   &  &  [km\,s$^{-1}$]   &  [km\,s$^{-1}$]   &   [km\,s$^{-1}$\,arcmin$^{-1}$]   &   [km\,s$^{-1}$]   &  [$^{\circ}$] & \\
		\hline     
				    & Linear rotation & $-$142.2$^{+1.8}_{-1.8}$ & 10.3$^{+1.6}_{-1.3}$ & $-$5.0$^{+1.6}_{-1.9}$ &  & 139.0$^{+17.4}_{-26.8}$ & ln$B_{\rm lin,flat}=1.7$ \\
			 All & Flat rotation & $-$142.4$^{+2.0}_{-2.0}$ & 11.2$^{+1.6}_{-1.4}$ &  & $-$7.1$^{+2.9}_{-3.0}$ & 135.6$^{+22.0}_{-29.8}$ & ln$B_{\rm rot,disp}=1.6$ \\
			 & Dispersion-only & $-$141.5$^{+1.9}_{-2.1}$ & 12.0$^{+1.7}_{-1.5}$ &  &  &  & \\           
			 \hline
		 & Linear rotation & $-$141.1$^{+2.0}_{-2.1}$ & 8.2$^{+2.0}_{-1.6}$ & $-$3.7$^{+1.9}_{-2.1}$ &  & 87$^{+53}_{-49}$ & ln$B_{\rm lin,flat}=0.5$ \\
		MR & Flat rotation & $-$140.6$^{+2.1}_{-2.2}$ & 8.7$^{+2.1}_{-1.7}$ &  & $-$4.3$^{+3.4}_{-3.1}$ & 76$^{+49}_{-44}$ & ln$B_{\rm rot,disp}=-1.8$ \\
		 & Dispersion-only & $-$140.9$^{+2.1}_{-2.1}$ & 9.1$^{+2.1}_{-1.6}$ &  &  &  & \\                 
		\hline
		& Linear rotation & $-$142.9$^{+3.3}_{-3.2}$ & 13.1$^{+2.9}_{-2.2}$ & $-$5.8$^{+2.5}_{-2.7}$ &  & 145$^{+21}_{-32}$ & ln$B_{\rm lin,flat}=-0.8$ \\
		MP & Flat rotation & $-$144.6$^{+3.3}_{-3.2}$ & 12.9$^{+2.9}_{-2.2}$ &  & $-$14.5$^{+6.3}_{-6.1}$ & 157$^{+15}_{-22}$ & ln$B_{\rm rot,disp}=0.6$ \\
		 & Dispersion-only & $-$142.3$^{+3.7}_{-3.6}$ & 15.4$^{+3.2}_{-2.5}$ &  &  &  & \\                 
		\hline 
	\end{tabular}
\end{table*}

\section{Discussion}
\label{sec:discussion}

In Fig.~\ref{fig:comp_gas} we compare the structural and kinematic properties of the HI and stellar component of Aquarius. 

\citet{iorio2017} found a weak velocity gradient in the HI gas, of amplitude not dissimilar to what we measure for the stars at comparable radii, but with a kinematic major-axis of P.A. $=77.3\pm15.2$ degrees (in their definition this implies receding velocities on the East side). This velocity gradient is misaligned and counter-rotating with the kinematic P.A. of the stellar component, as well as with the P.A. of the surface density maps of the stellar and HI components (see ellipses in Fig.~\ref{fig:comp_gas}). 
    
It is possible that the P.A. of the optical component is affected by the (small fraction) of bright young stars in Aquarius. On the other hand, \citet{iorio2017} note that the HI map is quite peculiar with iso-density contours that are not elliptical. In fact, judging from Fig.~\ref{fig:comp_gas}, the density map of the HI component appears to have a P.A. of $\sim$130-140 degrees and there appears to be HI missing in the S-E quadrant around that position angle \citep[see also][]{mcconnachie2006}. This might raise the question of whether, should HI have been present in this region, the kinematic P.A. of the HI component could be reconciled with the kinematic P.A. of the stars.

Nonetheless, it is clear that the HI gas and the RGB stars appear to counter-rotate, with the former having the most negative velocities on the West side, while the latter displaying them on the East side. In Sect.~\ref{mock} it was established that should the underlying kinematic properties of Aquarius stellar component be like the HI gas, there would be less than 1\% probability of measuring the observed misalignment between the HI and stellar kinematic major axes.  
    
In Fig.~\ref{fig:comp_gas} we compare directly the stellar (left-hand panel) and HI (middle panel) velocity fields. In order to make a pixel by pixel comparison of the velocity fields, we binned the Aquarius FORS2 members using the same pixel size of the HI map (1.5 arcsec). Then, we applied a Voronoi binning technique increasing the Poisson signal-to-noise of each bin to 3 ($\approx$ 9 star per Voronoi bin). The resultant velocity field for the stars is shown as colored-areas in the left-hand panel of Fig.~\ref{fig:comp_gas}. The presence of a velocity  gradient is obvious, approximately along the stellar major axis in agreement with the results obtained in Sec.~\ref{sec:kinematics}. The right-hand panel of Fig. 7 shows the pixel by pixel difference between the stellar (left-hand panel) and the HI velocity field (middle panel). The velocity difference is approximately $5$\,km\,s$^{-1}$ along the HI P.A. and reaches $\approx$ 10\,km\,s$^{-1}$ close to the East and North-West edges of the HI disc.

This phenomenon of counter- or misaligned rotation of two different components of a galaxy has been already observed in the Local Group dwarf galaxy NGC~6822 \citep{demers2006} and systems outside the Local Group.
The first event of counter-rotation was reported by \citet{bettoni1984} when studying the stellar and gas kinematics of six elliptical galaxies. It could be related to mergers with other galaxies that may have determined the internal evolution of the systems, to internal instabilities \citep[][]{evans1994} or to accretion of the gas, which should have produced star formation, so two different stellar populations can be differentiated \citep[][]{pizzella2004}.
    
Recently \citet{TStarkenburg2019} tried to understand the origin of such counter-rotation using a large sample of low mass galaxies ($M_{\star} \sim 10^{9} - 10^{10} M_{\odot}$) from the Illustris simulations. They found that only $\sim$1\% of their sample showed signs of star-gas counter-rotation at the present time, when considering discs and spheroids together. The origin of counter-rotation was ascribed to a significant episode of gas-loss followed by the acquisition of new gas with misaligned angular momentum. They identified two main mechanisms for the gas removal: internally induced by a strong feedback burst or environmentally induced by a fly-by passage with a large host causing the gas stripping. On the other hand they found no significant relation between the counter-rotation and the presence of a major merger event. 
Taking into account the extreme isolation at which Aquarius is found, the hypothesis of the internally induced counter-rotation seems appealing. In \citet{TStarkenburg2019}, galaxies exhibiting counter-rotation were predominantly found among dispersion dominated systems. Given that in general Local Group dwarf galaxies in the Aquarius stellar mass range are not rotation supported, it is possible that in this regime the overall fraction of galaxies that could have experienced events resulting in counter-rotation of gas and stars could be larger than the 1\% estimated in \citet{TStarkenburg2019}. 

\citet{iorio2017} find signs of a possible inflow/outflow of gas in Aquarius, as an extended region of HI emission along the minor axis not connected with the rotating HI disc. We postulate that in general the kinematics of the HI component in Aquarius might be dominated by recently accreted gas, while the RGB stars, which according to the \citet{cole2014} SFH are likely to be dominated by $\sim$8\,Gyr old stars, are tracing the kinematic properties as they were imprinted a much longer time ago along a different kinematic axis.  
    
\section{Summary and conclusions}
\label{sec:conclusions}

    We present an analysis of the kinematic and metallicity properties of the isolated Local Group dwarf galaxy Aquarius. The data-set consisted of VLT/FORS2 MXU spectroscopic observations in the region of the near-IR CaT for 53 individual targets. The spectra have a median SNR of 26 pix$^{-1}$ and led to the determination of l.o.s. velocities and [Fe/H] measurements with median uncertainties of $\pm$ 4.8 and 0.13 dex, respectively. Of the 53 individual stars observed, 45 are probable RGB stars that are members of Aquarius, which doubles the number of RGB stars with l.o.s. velocities and [Fe/H] measurements available in the literature for this galaxy. 
    
    The systemic velocity derived for Aquarius is $-142.2\pm 1.8$\,km\,s$^{-1}$, in agreement with prior determinations from samples of individual stars \citep{kirby2017} and also fully consistent with that of the HI component \citep{iorio2017}. 
    
    We find the internal kinematics of Aquarius to be best modelled by a combination of random motions (with l.o.s. velocity dispersion $=10.3_{-1.3}^{+1.6}$\,km\,s$^{-1}$) and linear rotation, with a velocity gradient of  $-5.0^{+1.6}_{-1.9}$\,km\,s$^{-1}$\,arcmin$^{-1}$ along an axis with P.A.$= 139^{+17}_{-27}$ degrees, broadly consistent (within 2$\sigma$) with the optical projected major axis of the galaxy. 
    
    On the other hand, the HI gas has a weak velocity gradient of comparable amplitude but along an axis with P.A.$=77.3\pm15.2$ degrees \citep{iorio2017}. According to the definitions used in this work, this implies counter-rotation of the stellar and HI component. 
    
    We have run a set of mock tests to better understand the rotational properties that can be derived from the FORS2 data. The results of these tests indicate that such misalignment is not the result of the characteristics of our FORS2 data (number statistics, coverage, measurement errors). A direct comparison of the stars and HI velocity fields lends further support to the detection of such counter-rotation.
    
    We speculate that the kinematics of the HI is dominated by recently accreted gas which is not tracing the kinematic properties of the RGB stars, the bulk of which are likely to have formed $\approx$8 Gyr ago \citep{cole2014} \citep[although the observed sample is likely to be biased towards younger stars;][]{Manning2017}. Possibly, this HI gas could simply be gas within Aquarius which was affected by particularly strong episodes of internal stellar feedback, rather than having been recently acquired from the intergalactic medium.

    Finally, we have characterized the metallicity properties of Aquarius. 
    The median metallicity ([Fe/H]$=-1.59 \pm 0.05$\,dex) indicates that it is a metal-poor galaxy, in agreement with the results from \citet{kirby2017}. We have analyzed the distribution of the metallicities as a function of radius, characterizing them through a running median and a Gaussian process regression; this has shown the presence of a very mild negative metallicity gradient, with the more metal-rich stars found in the innerparts of the galaxy. Should the presence of such a gradient be confirmed with larger data-sets, it would add to the number of isolated Local Group dwarf galaxies that display negative metallicity gradients and live in an environment where interactions with the large LG spirals cannot be invoked to explain these properties. 

\begin{acknowledgements}
The authors thank the referee for a thorough report that has helped improving the manuscript. L.H.M. acknowledges financial support from the State Agency for Research of the Spanish MCIU through the \lq Center of Excellence Severo Ochoa\rq award for the Instituto de Astrof\'{i}sica de Andaluc\'{i}a (SEV-2017-0709) and through the grants AYA2016-76682-C3 and BES-2017-082471. G.B. and S.T. acknowledge financial support through the grants (AEI/FEDER, UE) AYA2017-89076-P, AYA2014-56795-P, as well as by the Ministerio de Ciencia, Innovaci\'on y Universidades (MCIU), through the State Budget and by the Consejer\'\i a de Econom\'\i a, Industria, Comercio y Conocimiento of the Canary Islands Autonomous Community, through the Regional Budget. G.B. acknowledges financial support through the grant RYC-2012-11537. E.S. gratefully acknowledges funding by the Emmy Noether program from the Deutsche Forschungsgemeinschaft (DFG).

This research has made use of NASA's Astrophysics Data System, IRAF and Python, specifically with Astropy, (\nolinkurl{http://www.astropy.org}), a community-developed core Python package for Astronomy \citep{astropy:2013, astropy:2018}, Scipy, Matplotlib \citep{Hunter:2007}, Scikit-learn \citep{scikit-learn} and PyMultinest \citep{pymultinest2014}.
\end{acknowledgements}

%
   \bibliographystyle{aa} 
   \bibliography{aquarius_v2_referee.bbl} 

%






   
%
%
%


\begin{appendix} 
\section{Output tables}

\begin{table*}
    \caption{Observing log of VLT/FORS2 MXU observations of RGB targets along 
	the line-of-sight to the Aquarius dwarf galaxy. The columns of the table from left to right 
	indicate: the ID of each observation; the coordinates of the exposure RA/DEC; the date in 
	which the observations were made; the total exposure time of each observation; 
	the mean airmass during the observation; the average seeing during the exposure in 
	arcsec; and the ESO Observation Block (OB) fulfillment grades$^{(*)}$.}
	\label{table:log}      
	\centering          
	\begin{tabular}{c c c c c c c c c c}    
		\hline \hline
		Mask & Obs. ID  & RA & DEC & Observation date & Exp. time & Airmass & Seeing & Grade$^{(*)}$\\ 
		 &  & (J2000) & (J2000) & (UT) & (sec) &  & (arcsec) & \\
		\hline   
Aquarius0 & 972757 & 23:42:40.07 & -12:51:13.32 & 2013-07-17 / 06:18:16 & 3400 & 1.06 & 0.85 & A\\     
 & 972765 & 23:42:40.04 & -12:51:13.18 & 2013-07-07 / 04:46:35 & 3400 & 1.08 & 0.87 & A\\
 & 972768 & 23:42:40.05 & -12:51:13.21 & 2013-07-07 / 06:06:06 & 3400 & 1.03 & 1.04 & A\\
 & 972771 & 23:42:40.06 & -12:51:13.25 & 2013-07-07 / 08:07:34 & 3400 & 1.20 & 1.20 & A\\
 & 972774 & 23:42:40.04 & -12:51:13.28 & 2013-07-10 / 04:51:34 & 3400 & 1.06 & 1.03 & A\\
 & 972777 & 23:42:40.04 & -12:51:13.28 & 2013-07-10 / 07:25:27 & 3400 & 1.12 & 0.85 & A\\
 & 972780 & 23:42:40.13 & -12:51:13.36 & 2013-07-17 / 07:30:43 & 3400 & 1.21 & 0.79 & A\\
 & 972783 & 23:42:40.03 & -12:51:13.28 & 2013-08-03 / 05:12:07 & 3400 & 1.06 & 0.79 & A\\
 & 972786 & 23:42:40.10 & -12:51:13.32 & 2013-08-29 / 03:38:59 & 3400 & 1.07 & 1.27 & A\\
 & 972789 & 23:42:40.09 & -12:51:13.28 & 2013-08-29 / 04:50:51 & 3400 & 1.24 & 0.9 & B\\
 \hline
Aquarius1 & 972792 & 23:42:58.14 & -12:50:53.27 & 2013-09-01 / 01:35:02 & 3800 & 1.05 & 1.32 & B\\
 & 972800 & 23:42:58.14 & -12:50:53.27 & 2013-09-01 / 02:46:25 & 3400 & 1.04 & 1.10 & B\\
 & 972803 & 23:42:58.10 & -12:50:53.30 & 2013-09-01 / 03:52:10 & 3800 & 1.12 & 1.18 & B\\
 & 972806 & 23:42:58.15 & -12:50:53.05 & 2013-06-06 / 07:41:52 & 3400 & 1.03 & 0.93 & A\\
 & 972809 & 23:42:57.84 & -12:50:52.80 & 2013-06-07 / 06:36:19 & 3400 & 1.09 & 0.60 & A\\
 & 972812 & 23:42:57.84 & -12:50:52.80 & 2013-06-07 / 07:37:59 & 3400 & 1.03 & 0.54 & A\\
 & 972815 & 23:42:58.20 & -12:50:53.02 & 2013-06-07 / 08:49:10 & 3400 & 1.05 & 0.53 & A\\
 & 972818 & 23:42:58.10 & -12:50:53.02 & 2013-09-03 / 01:05:20 & 3900 & 1.07 & 1.59 & B\\
 & 972821 & 23:42:58.10 & -12:50:53.02 & 2013-09-03 / 02:18:48 & 3400 & 1.03 & 1.43 & B\\
 & 972824 & 23:42:58.10 & -12:50:53.02 & 2013-09-03 / 03:24:23 & 3400 & 1.08 & 1.33 & B\\
		\hline        
	\end{tabular}\\
	\tablefoot{$^{(*)}$ESO OB fulfillment Grades: A) Fully within constraints - OB completed; B) Mostly within constraints, some constraint is  $10\%$ violated - OB completed.}
\end{table*}

\begin{table*}
	\caption{Summary of the results for the 53 target stars in the line-of-sight of Aquarius. The columns represent, from left to right: name, RA-DEC of the targets; derived heliocentric line-of-sight velocity and its error; S/N ratio; V and I magnitude of the stars and its error (obtained from \citet{mcconnachie2006}), and the metallicity with its error. In the last column "K" indicates the targets in common with \citet{kirby2017}, "rep" indicates the stars that have been measured twice (the results have been combined for the two repeated stars, note that \textit{aqu1c1star13} and \textit{aqu1c1star17} are missing), "C" indicates the stars containing strong CN bands; "N" indicate the stars excluded from the analysis (non-members).}
	\label{tabla:stars_aqu}
	\centering
	\begin{tabular}{c c c c c c c c c}
		\hline \hline
		Star & RA (deg) & DEC (deg) & $v_{\rm hel} \pm \delta$ $v_{\rm hel}$ & S/N & I $\pm \delta$I & V $\pm \delta$V & [Fe/H] $\pm \delta$[Fe/H] & Com\\
		 & (J2000) & (J2000) & (\,km\,s$^{-1}$) & (pxl$^{-1}$) &  &  & (dex) & \\
		\hline 
		aqu0c1star1 & $311.7103$ & $-12.85993$ & $77.6 \pm 4.9$ & $49.0$ & $20.535 \pm 0.003$ & $21.798 \pm 0.007$ & $-1.88 \pm 0.09$ & N\\
		aqu0c1star2 & $311.7043$ & $-12.83635$ & $-136.4 \pm 5.2$ & $27.4$ & $21.321 \pm 0.005$ & $22.592 \pm 0.012$ & $-1.00 \pm 0.16$ & \\
		aqu0c1star3 & $311.7062$ & $-12.83348$ & $-141.4 \pm 5.4$ & $37.4$ & $20.977 \pm 0.004$ & $22.277 \pm 0.010$ & $-1.79 \pm 0.09$ & \\
		aqu0c1star4 & $311.7089$ & $-12.83242$ & $-65.2 \pm 4.5$ & $41.7$ & $20.606 \pm 0.003$ & $22.205 \pm 0.009$ & $-2.09 \pm 0.08$ & N\\
		aqu0c1star5 & $311.7146$ & $-12.83425$ & $-145.0 \pm 5.2$ & $54.4$ & $20.612 \pm 0.003$ & $21.999 \pm 0.008$ & $-1.59 \pm 0.08$ & \\
		aqu0c1star6 & $311.7189$ & $-12.83611$ & $-154.9 \pm 4.2$ & $37.5$ & $20.835 \pm 0.003$ & $22.250 \pm 0.010$ & $-1.26 \pm 0.08$ &\\
		aqu0c1star7 & $311.7242$ & $-12.84258$ & $-143.7 \pm 3.6$ & $20.7$ & $21.559 \pm 0.005$ & $22.674 \pm 0.013$ & $-0.99 \pm 0.16$ & \\
		aqu0c1star8 & $311.7264$ & $-12.83883$ & $-130.9 \pm 8.5$ & $22.8$ & $21.412 \pm 0.005$ & $22.685 \pm 0.013$ & $-1.24 \pm 0.16$ & \\
		aqu0c1star9 & $311.7368$ & $-12.85841$ & $-133.8 \pm 4.3$ & $42.6$ & $20.903 \pm 0.004$ & $22.294 \pm 0.010$ & $-1.51 \pm 0.09$ & \\
		aqu0c1star10 & $311.7401$ & $-12.85359$ & $-165.9 \pm 8.6$ & $12.2$ & $20.923 \pm 0.004$ & $22.351 \pm 0.010$ & $-1.44 \pm 0.08$ & C\\
		aqu0c1star11 & $311.7488$ & $-12.86734$ & $-177.0 \pm 9.0$ & $17.2$ & $21.566 \pm 0.005$ & $22.721 \pm 0.014$ & $-2.09 \pm 0.15$ & rep/K\\
		aqu0c1star12 & $311.7316$ & $-12.81242$ & $-148.1 \pm 5.1$ & $32.5$ & $21.226 \pm 0.004$ & $22.504 \pm 0.011$& $-1.78 \pm 0.10$ & \\
		aqu0c1star13 & $311.7470$ & $-12.84501$ & $-154.9 \pm 4.9$ & $28.5$ & $21.297 \pm 0.004$ & $22.542 \pm 0.012$ & $-1.48 \pm 0.13$ & \\
		aqu0c1star14 & $311.7436$ & $-12.82085$ & $-143.8 \pm 4.3$ & $29.7$ & $21.451 \pm 0.005$ & $22.655 \pm 0.013$ & $-1.77 \pm 0.16$ & \\
		aqu0c1star15 & $311.7558$ & $-12.84592$ & $-173.1 \pm 15.6$ & $23.3$ & $21.294 \pm 0.004$ & $22.608 \pm 0.012$ & $-2.13 \pm 0.09$ & \\
		aqu0c1star16 & $311.7645$ & $-12.85741$ & $-149.6 \pm 4.7$ & $36.2$ & $20.696 \pm 0.003$ & $22.169 \pm 0.009$ & $-1.36 \pm 0.08$ & rep\\
		aqu0c1star17 & $311.7687$ & $-12.85716$ & $-154.8 \pm 8.8$ & $22.3$ & $21.495 \pm 0.005$ & $22.694 \pm 0.013$ & $-1.74 \pm 0.18$ & \\
		\hline
		aqu0c2star1 & $311.6577$ & $-12.86324$ & $-76.5 \pm 4.4$ & $47.3$ & $20.787 \pm 0.003$ & $22.174 \pm 0.009$ & $-1.96 \pm 0.11$ & N\\
		aqu0c2star2 & $311.6774$ & $-12.89136$ & $-278.8 \pm 4.8$ & $30.8$ & $20.967 \pm 0.004$ & $22.230 \pm 0.010$ & $-2.17 \pm 0.12$ & N \\
		aqu0c2star3 & $311.6665$ & $-12.83773$ & $-119.5 \pm 3.9$ & $23.8$ & $21.410 \pm 0.005$ & $22.623 \pm 0.012$ & $-1.63 \pm 0.15$ & K\\
		aqu0c2star4 & $311.6754$ & $-12.85405$ & $-135.0 \pm 4.4$ & $18.7$ & $21.612 \pm 0.005$ & $22.697 \pm 0.013$ & $-1.43 \pm 0.13$ &\\
		aqu0c2star5 & $311.6755$ & $-12.84307$ & $-138.8 \pm 5.4$ & $24.8$ & $21.241 \pm 0.004$ & $22.475 \pm 0.011$ & $-1.42 \pm 0.14$ & \\
		aqu0c2star6 & $311.6865$ & $-12.85343$ & $-140.2 \pm 5.1$ & $25.9$ & $21.436 \pm 0.005$ & $22.547 \pm 0.012$ & $-1.64 \pm 0.17$ & \\
		aqu0c2star7 & $311.6910$ & $-12.85157$ & $-124.4 \pm 3.8$ & $31.9$ & $21.214 \pm 0.004$ & $22.506 \pm 0.012$ & $-1.32 \pm 0.12$ & K\\
		aqu0c2star8 & $311.6935$ & $-12.84941$ & $-144.1 \pm 4.5$ & $25.3$ & $21.504 \pm 0.005$ & $22.560 \pm 0.012$ & $-1.34 \pm 0.17$ & \\
		aqu0c2star9 & $311.7057$ & $-12.87361$ & $94.0 \pm 4.8$ & $33.9$ & $21.162 \pm 0.004$ & $22.575 \pm 0.012$ & $-1.51 \pm 0.11$ & N\\
		
		\hline\hline
		
		aqu1c1star1 & $311.7108$ & $-12.84064$ & $-149.9 \pm 4.9$ & $22.7$ & $21.458 \pm 0.005$ & $22.829 \pm 0.015$ & $-1.35 \pm 0.17$ & \\
		aqu1c1star2 & $311.7135$ & $-12.82870$ & $-138.4 \pm 4.4$ & $23.1$ & $21.214 \pm 0.004$ & $22.462 \pm 0.011$ & $-1.44 \pm 0.18$ & \\
		aqu1c1star3 & $311.7175$ & $-12.85540$ & $-140.2 \pm 4.4$ & $26.9$ & $21.376 \pm 0.005$ & $22.667 \pm 0.013$ & $-1.20 \pm 0.16$ & \\
		aqu1c1star4 & $311.7204$ & $-12.84027$ & $-141.8 \pm 19.9$ & $6.5$ & $20.336 \pm 0.003$ & $22.411 \pm 0.011$ & $-1.95 \pm 0.05$ & C\\
		aqu1c1star5 & $311.7244$ & $-12.85543$ & $-153.6 \pm 3.8$ & $30.7$ & $21.165 \pm 0.004$ & $22.402 \pm 0.011$ & $-1.39 \pm 0.12$ & \\
		aqu1c1star6 & $311.7270$ & $-12.84309$ & $-136.6 \pm 4.3$ & $24.5$ & $21.292 \pm 0.004$ & $22.372 \pm 0.011$ & $-1.39 \pm 0.14$ & \\
		aqu1c1star7 & $311.7297$ & $-12.84016$ & $-116.5 \pm 3.1$ & $30.8$ & $21.047 \pm 0.004$ & $22.214 \pm 0.009$ & $-1.99 \pm 0.09$ & \\
		aqu1c1star8 & $311.7342$ & $-12.83716$ & $-131.6 \pm 2.7$ & $26.5$ & $21.300 \pm 0.004$ & $22.600 \pm 0.012$ & $-1.36 \pm 0.15$ & \\
		aqu1c1star9 & $311.7367$ & $-12.84172$ & $-137.5 \pm 7.5$ & $20.7$ & $21.574 \pm 0.004$ & $22.754 \pm 0.014$ & $-1.27 \pm 0.18$ & \\
		aqu1c1star10 & $311.7399$ & $-12.83490$ & $-143.2 \pm 4.8$ & $29.6$ & $21.180 \pm 0.004$ & $22.551 \pm 0.012$ & $-1.70 \pm 0.13$ & K\\
		aqu1c1star11 & $311.7429$ & $-12.83229$ & $-141.8 \pm 5.8$ & $23.9$ & $21.442 \pm 0.005$ & $22.652 \pm 0.013$ & $-1.79 \pm 0.17$ & \\
		aqu1c1star12 & $311.7460$ & $-12.86016$ & $-150.0 \pm 3.9$ & $24.9$ & $21.244 \pm 0.004$ & $22.485 \pm 0.011$ & $-1.76 \pm 0.17$ & \\
		aqu1c1star14 & $311.7542$ & $-12.88221$ & $-160.9 \pm 6.2$ & $25.9$ & $21.042 \pm 0.004$ & $22.325 \pm 0.010$ & $-1.67 \pm 0.12$ & \\
		aqu1c1star15 & $311.7573$ & $-12.83783$ & $-321.5 \pm 16.9$ & $22.3$ & $21.566 \pm 0.004$ & $22.667 \pm 0.013$ & $-1.96 \pm 0.17$ & N\\
		aqu1c1star16 & $311.7607$ & $-12.85800$ & $-74.3 \pm 5.2$ & $53.1$ & $20.385 \pm 0.003$ & $20.896 \pm 0.004$ & $-2.90 \pm 0.10$ & N\\
		aqu1c1star18 & $311.7675$ & $-12.85755$ & $-148.8 \pm 5.1$ & $39.9$ & $20.703 \pm 0.003$ & $22.030 \pm 0.008$ & $-1.75 \pm 0.11$ & \\
		\hline 
		aqu1c2star1 & $311.6636$ & $-12.84180$ & $-139.5 \pm 13.8$ & $7.4$ & $20.448 \pm 0.003$ & $22.237 \pm 0.009$ & $-1.45 \pm 0.06$ & C \\
		aqu1c2star2 & $311.6671$ & $-12.86534$ & $-92.0 \pm 4.7$ & $41.2$ & $20.668 \pm 0.003$ & $22.261 \pm 0.010$ & $-1.91 \pm 0.09$ & N \\
		aqu1c2star3 & $311.6708$ & $-12.85654$ & $-137.2 \pm 4.3$ & $19.3$ & $21.666 \pm 0.006$ & $22.790 \pm 0.014$ & $-1.66 \pm 0.21$ & K\\
		aqu1c2star4 & $311.6755$ & $-12.84781$ & $-121.9 \pm 4.3$ & $25.5$ & $21.338 \pm 0.005$ & $22.612 \pm 0.012$ & $-1.56 \pm 0.13$ & \\
		aqu1c2star5 & $311.6796$ & $-12.83744$ & $-157.5 \pm 4.7$ & $22.9$ & $21.398 \pm 0.005$ & $22.476 \pm 0.011$ & $-1.75 \pm 0.16$ & \\
		aqu1c2star6 & $311.6843$ & $-12.84784$ & $-123.4 \pm 11.4$ & $10.9$ & $21.549 \pm 0.005$ & $22.784 \pm 0.014$ & $-1.91 \pm 0.13$ & C \\
		aqu1c2star7 & $311.6869$ & $-12.83789$ & $-131.4 \pm 5.3$ & $23.1$ & $21.121 \pm 0.004$ & $22.311 \pm 0.010$ & $-2.09 \pm 0.12$ & \\
		aqu1c2star8 & $311.6924$ & $-12.83748$ & $-139.2 \pm 5.0$ & $22.4$ & $21.572 \pm 0.005$ & $22.746 \pm 0.014$ & $-1.82 \pm 0.12$ & \\
		aqu1c2star9 & $311.6960$ & $-12.85198$ & $-151.9 \pm 3.5$ & $37.0$ & $21.012 \pm 0.004$ & $22.295 \pm 0.010$ & $-2.27 \pm 0.10$ & \\
		aqu1c2star10 & $311.7009$ & $-12.84057$ & $-141.3 \pm 2.2$ & $43.8$ & $20.588 \pm 0.003$ & $22.048 \pm 0.008$ & $-1.56 \pm 0.08$ & \\
		aqu1c2star11 & $311.7038$ & $-12.82979$ & $-110.6 \pm 3.5$ & $26.7$ & $21.233 \pm 0.004$ & $22.499 \pm 0.011$ & $-1.63 \pm 0.12$ & \\
		\hline
	\end{tabular}
\end{table*}

\end{appendix}

\end{document}